\documentclass[fleqn,usenatbib]{mnras}

\usepackage{newtxtext,newtxmath}
\usepackage[T1]{fontenc}
\usepackage{xcolor}

\usepackage{soul}
\let\VANthebibliography\thebibliography
\def\thebibliography{\DeclareRobustCommand{\VAN}[3]{##3}\VANthebibliography}

\usepackage{graphicx}	% Including figure files
\usepackage{amsmath}	% Advanced maths commands
\usepackage{siunitx}
\usepackage{comment}
\usepackage{float}
\usepackage{placeins}
\usepackage{pifont}
\usepackage{footnote}

\newcommand{\cmark}{\ding{51}}%
\newcommand{\xmark}{\ding{55}}%
\newcommand{\mobse}{\textsc{mobse}}

\title[Clustering of binary black hole mergers]{Clustering of binary black hole mergers:\\ a detailed analysis of the EAGLE+MOBSE simulation}

\author[Peron M. et al.]{%
Matteo Peron,$^{1,2,3}$\thanks{E-mail: matteo.peron@unipr.it}
Sarah Libanore,$^{4,2,3}$
Andrea Ravenni,$^{2,3}$
\newauthor
Michele Liguori,$^{2,3,5,6}$
and Maria Celeste Artale$^{2,3,7}$
\\
$^{1}$ Department of Mathematical, Physical and Computer Sciences, University of Parma, Parco Area delle Scienze 7/A, 43124, Parma, Italy\\
$^{2}$ Dipartimento di Fisica e Astronomia G. Galilei, Universit\`{a} degli Studi di Padova  Via Marzolo 8, 35131 Padova, Italy\\
$^{3}$ INFN, Sezione di Padova, Via Marzolo 8, I$-$35131, Padova, Italy\\
$^{4}$ Department of Physics, Ben-Gurion University of the Negev, Be'er Sheva 84105, Israel\\
$^{5}$ Dipartimento di Fisica, Universit\`a degli Studi di Trento, Via Sommarive 14, I-38123 Povo (TN), Italy \\
$^{6}$INAF - Osservatorio Astronomico di Padova, Vicolo dell'Osservatorio 5, I-35122 Padova, Italy\\
$^{7}$ Departamento de Ciencias F\'isicas, Universidad Andr\'es Bello Fernandez Concha 700, Las Condes, Santiago, Chile}

\date{Accepted XXX. Received YYY; in original form ZZZ}

\pubyear{2023}

\begin{document}

\DeclareRobustCommand{\VAN}[3]{#2}

%% Maths

\newcommand{\p}{\partial}
\newcommand{\pd}[2]{\frac{\partial{#1}}{\partial{#2}}}
\newcommand{\diff}{\mathop{}\!\mathrm{d}}
\newcommand\Diff[1]{\mathop{}\!\mathrm{d^#1}}
\newcommand{\td}[2]{\frac{\diff #1}{\diff #2}}
%-------------------------------------------------
\newcommand{\threej}[6]{ \begin{pmatrix}
    #1 & #2 & #3 \\
    #4 & #5 & #6
\end{pmatrix}}
%-------------------------------------------------
\newcommand{\sixj}[6]{ \begin{Bmatrix}
    #1 & #2 & #3 \\
    #4 & #5 & #6
\end{Bmatrix}}
%-------------------------------------------------
%\renewcommand{\vec}[1]{\mathbf{#1}}
\renewcommand{\vec}[1]{{\bm#1}}
\newcommand{\vers}[1]{\hat{\vec{#1}}}
\newcommand{\Var}[1]{\operatorname{Var}(#1)}
\newcommand{\Cov}[2]{\mathbf{Cov}(#1,#2)}
\newcommand{\jprob}{{;\,}}
\newcommand{\intthreek}[1]{\int \frac{\Diff{3} \vec{#1}}{(2\pi)^3}}
\newcommand{\bgw}{b_{\rm m}}
%-------------------------------------------------
%paper specific
\newcommand{\Mchirp}{\mathcal{M}_{c}}%\bullet\bullet}}
\newcommand{\mrem}{m}

%==================================================================================================

%% Experiments and acronyms
\newcommand{\Planck}{\textit{Planck}\xspace}
\newcommand{\WMAP}{WMAP\xspace}
\newcommand{\COBEFIRAS}{COBE/FIRAS}
\newcommand{\CLASS}{\texttt{CLASS}\xspace}
\newcommand{\MontePython}{Monte-Python\xspace}
\newcommand{\CMBXIII}{CMB13\xspace}
\newcommand{\CMBXV}{CMB15\xspace}
\newcommand{\SDSSDRVII}{SDSS~DR7\xspace}
\newcommand{\WiggleZ}{WiggleZ\xspace}
\newcommand{\PlanckLens}{{\Planck}Lens\xspace}
\newcommand{\CFHTLenS}{CFHTLenS\xspace}
\newcommand{\CFHTLens}{\CFHTLenS\xspace}
%==================================================================================================

%% Family names
\newcommand{\Thomson}{Thomson\xspace}
\newcommand{\Rayleigh}{Rayleigh\xspace}
\newcommand{\Friedmann}{Friedmann\xspace}
\newcommand{\Lemaitre}{Lema\^itre\xspace}
\newcommand{\Robertson}{Robertson\xspace}
\newcommand{\Walker}{Walker\xspace}
\newcommand{\Bianchi}{Bianchi\xspace}
\newcommand{\DeSitter}{de Sitter\xspace}
\newcommand{\Mukhanov}{Mukhanov\xspace}
\newcommand{\Hankel}{Hankel\xspace}
\newcommand{\BunchDavies}{Bunch-Davies\xspace}
\newcommand{\Liouville}{Liouville\xspace}
\newcommand{\SW}{Sachs-Wolfe\xspace}
\newcommand{\Lightman}{Lightman\xspace}
\newcommand{\Gould}{Gould\xspace}
\newcommand{\Heaviside}{Heaviside\xspace}
%==================================================================================================

%% 		COSMOLOGICAL STUFF
\newcommand{\igmn}{g^{\mu\nu}}
\newcommand{\gmn}{g_{\mu\nu}}
\DeclareRobustCommand{\munu}{{\mu\nu}}
\newcommand{\fnl}{f_\text{NL}}
\newcommand{\fNL}{\fnl}

\newcommand{\Tref}{T_\text{ref}}
\newcommand{\Tg}{T_\gamma}

\newcommand{\vb}{v_\text{b}}

\newcommand{\sigmaT}{\sigma_\text{T}}

\newcommand{\SZ}{Sunyaev-Zeldovich\xspace}
\newcommand{\PPS}{PPS\xspace}
\newcommand{\powerspectrum}{power-spectrum\xspace}
\newcommand{\loglikelihood}{log-likelihood\xspace}
\newcommand{\zetak}{\zeta_{\vec{k}}}
\newcommand{\As}{A_\text{s}}
\newcommand{\kp}{k_\text{p}}
\newcommand{\kl}{k_\text{l}}
\newcommand{\ks}{k_\text{s}}

%--------------------------------------------
%% ell m stuff
\newcommand{\alm}{a_{\ell m}}
\newcommand{\alpmp}{a_{\ell' m'}}
\newcommand{\Ylm}{Y_{\ell m}}
\newcommand{\Ylpmp}{Y_{\ell' m'}}
\newcommand{\Cl}{C_\ell}
\newcommand{\Clp}{C_{\ell'}}
\renewcommand{\lll}{{\ell_1 \ell_2 \ell_3}}
\newcommand{\lllp}{{\ell_4 \ell_5 \ell_6}}
\newcommand{\mmm}{{m_1 m_2 m_3}}
\newcommand{\mmmp}{{m_4 m_5 m_6}}
\newcommand{\lmax}{{\ell_\text{max}}}
\newcommand{\lu}{\ell_1}
\newcommand{\ld}{\ell_2}
\newcommand{\lt}{\ell_3}

%--------------------------------------------
%% Compton scattering related stuff
\newcommand{\Te}{T_\text{e}}
\newcommand{\me}{m_\text{e}}
\newcommand{\nume}{N_\text{e}}
\newcommand{\fe}{f_\text{e}}
\newcommand{\Ee}{E_\text{e}}
\newcommand{\barEe}{\bar{E}_\text{e}}
\newcommand{\wmin}{w_\text{min}}
\newcommand{\wmax}{w_\text{max}}
\newcommand{\wt}{w_\text{t}}
\newcommand{\wc}{w_\text{c}}
\newcommand{\mue}{\mu_{\text{e}0}}
\newcommand{\muuno}{\mu_{01}}
\newcommand{\mutwo}{\mu_{02}}
\newcommand{\phie}{\phi_\text{e}}
\newcommand{\phione}{\phi_1}
\newcommand{\phitwo}{\phi_2}
\newcommand{\wcusp}{w_2^\text{cusp}}

%==================================================================================================

%% Comments and review
\newcommand{\MP}[1]{\textcolor{red}{[{{\bf MP:} #1}]}}
\newcommand{\MPcut}[1]{\textcolor{red}{[{\bf MP:} \st{#1}]}}
\newcommand{\SL}[1]{\textcolor{orange}{[{{\bf SL:} #1}]}}
\newcommand{\SLcut}[1]{\textcolor{orange}{[{\bf SL:} \st{#1}]}}

\newcommand{\ML}[1]{\textcolor{teal}{[{{\bf ML:} #1}]}}

\DeclareRobustCommand{\red}[1]{{\leavevmode\color{red}#1}}
\DeclareRobustCommand{\grey}[1]{{\leavevmode\color{grey}#1}}

\DeclareRobustCommand{\ARc}[1]{ \textcolor{olive}{[AR: #1]} }
\DeclareRobustCommand{\ARC}[1]{\ARc{#1}}
\DeclareRobustCommand{\AR}[1]{{\leavevmode\color{olive}#1}}

\DeclareRobustCommand{\MRc}[1]{{\leavevmode\color{teal}[MR: #1]} }
\DeclareRobustCommand{\MRC}[1]{\MRc{#1}}
\DeclareRobustCommand{\MR}[1]{{\leavevmode\color{teal}#1}}

%\newcommand{\prd}{Physical Review D}
%\newcommand{\aap}{Astronomy \& Astrophysics}
%\newcommand{\apj}{The Astrophysical Journal}
%\newcommand{\apjs}{The Astrophysical Journal Supplement}
%\newcommand{\mnras}{Monthly Notices of the Royal Astronomical Society}
%\newcommand{\physrep}{Physics Reports}
%\newcommand{\jcap}{Journal of Cosmology and Astroparticle Physics}

%-------------------------------------------------

\newcommand{\eg}{\textit{e.g.,}\xspace}
\newcommand{\ie}{\textit{i.e.,}\xspace}
%-------------------------------------------------

\hyphenation{an-i-sot-ro-pies}
\hyphenation{an-i-sot-ro-py}

\label{firstpage}
\pagerange{\pageref{firstpage}--\pageref{lastpage}}
\maketitle

\begin{abstract}
We perform a detailed study of the cosmological bias of gravitational gave (GW) events produced by binary black hole mergers (BBHM). We start from a BBHM distribution modeled inside the \textsc{eagle} hydrodyamical simulation using the population synthesis code \mobse{}.
We then compare our findings with predictions from different Halo Occupation Distribution (HOD) prescriptions and find overall agreement, provided that the modeled properties of host galaxies and halos in the semi-analytical treatment match those in the simulations. By highlighting the sources of these discrepancies, we provide the stepping stone to build future more robust models that prevent the shortcoming of both simulation-based and analytical models.
Finally, we train a neural network to build a simulation-based HOD and perform feature importance analysis to gain intuition on which host halo/galaxy parameters are the most relevant in determining the actual distribution and power spectrum of BBHM. We find that the distribution of BBHM in a galaxy does not only depend on its size, star formation rate and metallicity, but also by its kinetic state.
\end{abstract}

\begin{keywords}
Bias -- Gravitational waves -- Large scale structures -- Cosmological simulations
\end{keywords}

%%%%%%%%%%%%%%%%%%%%%%%%%%%%%%%%%%%%%%%%%%%%%%%%%%

%%%%%%%%%%%%%%%%% BODY OF PAPER %%%%%%%%%%%%%%%%%%

\section{Introduction}

Future GW interferometers, such as the Einstein Telescope (ET) \citep{Punturo_2010,ET_2012,ET_2020} and Cosmic Explorer (CE) \citep{CE_2021}, are expected to detect $\sim 10^4$-$10^5$ merger events per year. In this regime, it will be possible to study the spatial distribution and clustering properties of GW sources in similar fashion as done for galaxies. These GW catalogues will contain a significantly smaller number of objects with respect to galaxy surveys and will have relatively poor sky localization, making the smallest scales unaccessible. 

Nevertheless, at least for binary black hole mergers, they will probe very large cosmic volumes, since they will be able to collect events from almost the full sky, with so high sensitivity that their horizon will reach very high redshifts ($z \sim 10$). Therefore, it will be possible to use them for high precision measurements of cosmological parameters, either by considering their angular power spectrum \citep{namikawa_2016,Libanore:2020fim,namikawa2021,Libanore:2021jqv,Mukherjee:2020hyn,Zhu:2021aat,Yang:2022uye}, or cross-correlating their distribution with that of other tracers \citep{scelfo2018,Scelfo:2020jyw,Balaudo:2022znx,Mukherjee:2022afz,Scelfo:2022lsx,Libanoreinprep}. Note that GW surveys do not allow for a direct measurement of the redshift at which the merger took place: thus, a tomographic analysis of the sources distribution has either to assume the background cosmology (as it is done e.g. for LIGO-Virgo-KAGRA catalogues~\citealt{ligo2016,LIGOScientific:2018mvr,ligo03,ligo03A}), to use cross-correlations or other statistical techniques (e.g., in~\citealt{delpozzo_2012, Scelfo_2022,Mukherjee:2020hyn}), or to rely only on luminosity distance (adopted, e.g., in~\citealt{namikawa_2016,Libanore:2020fim}) as a radial position indicator; all these approaches have been so far considered in the literature and have their specific advantages and drawbacks. Methodologies that infer the redshift and combine it with the luminosity distance of GW events also allow for interesting generalizations of the Alcock-Paczynski test \citep{Mukherjee:2018ebj}. Finally, it has been pointed out that GW spatial clustering could be used to detect the presence of Primordial Black Hole (PBH) merger events in a catalogue \citep{raccanelli_2016,scelfo2018,Scelfo:2020jyw,Libanoreinprep}. 

Since they entail the study of GW sources as Large Scale Structure (LSS) tracers, a key ingredient in all these applications is a detailed understanding of the GW cosmological bias properties, defining how the sources cluster with respect to the underlying dark matter (DM) field. 
The bias $\bgw$ of astrophysical merger events is linked with the underlying bias of both their host galaxies and DM halos. Based on how this link is modeled, a plethora of different prescriptions for $\bgw$ has been proposed in the literature. They range from assuming linear dependency on the host galaxy bias $b_{\rm g}$ ~\citep{Oguri:2016dgk}, to  redshift-dependent power laws (e.g.,~\citep{Mukherjee:2019qmm,Mukherjee:2020hyn,Vijaykumar:2020pzn,Libanore:2021jqv,Canas-Herrera:2021qxs}), to more refined parameterizations, accounting for the properties of the binary and/or its host galaxy, e.g.,~\citep{Mukherjee:2019oma,Calore:2020bpd,Bellomo:2021mer, Boco_2019,Scelfo:2020jyw,Libanore:2020fim,Capurri:2021zli}. 

This significant variety in bias models and assumptions is an issue that can propagate to generate systematic and statistical uncertainties in  cosmological studies of GW clustering. In this work, our main goal is to address and clarify this issue, focusing on a detailed study of the cosmological bias of black hole merger events. In order to capture the complexity of the binary formation process as accurately as possible, we rely on state-of-the-art results, obtained by combining binary black hole catalogues from population synthesis models and the outputs of hydrodynamical, cosmological simulations. More specifically, our dataset is based on the {\sc eagle}  cosmological, hydrodynamical simulation \citep{schaye2015} populated with black hole (BH) binaries through the \mobse{} population synthesis code \citep{Mapelli2017,Giacobbo:2017qhh} and evolved to get the final distribution of binary mergers, as described by~\citet{Mapelli2017,Mapelli2018,artale_2019,artale2020} (hereafter, A19, A20). We use these data to build models and predictions for the bias of Binary Black Hole Mergers (BBHM), in different ways. First, we implement a suitable estimator of the merger power spectrum and directly measure the bias from the simulations; second, we use a Halo Occupation Distribution (HOD) approach to build a semi-analytical model of the bias (following \citealt{Libanore:2021jqv}), based on the measured abundance of mergers in the simulations, as a function of the galaxy stellar mass and star formation rate. After testing the consistency of the results for internal validation purposes, we also compare them with the GW bias model of \citet{Scelfo:2020jyw} and we make an accurate assessment of how the starting assumptions in the different methodologies impact the final bias prediction.
Finally, we use the input simulations to train a neural network (NN) to map the relation between the number of GW sources and parameters describing the properties of their galaxy or DM halo hosts. The main purpose in this case is to understand whether the general models used in semi-analytical approaches capture all the relevant Physics or could be improved by including extra parameters.

The paper is organized as follows. Section~\ref{sec:sims} contains a summary description of the simulations used for our analysis; in Section~\ref{sec:bias} we briefly review some key aspects of the bias model that we consider in this work; in Section~\ref{sec:method} we describe our bias estimator, discuss the formulation and calibration of semi-analytical models for GW bias and give details on the implementation and architecture of our neural network. Results and conclusions can be found in Sections~\ref{sec:results} and~\ref{sec:conclusion}.

The self-consistent algorithm we developed to estimate the (more than tree-level) power spectrum and analyse the clustering properties of the {\sc eagle}+\mobse{} data-set is publicly available at \url{https://github.com/MatPeron/powerbias}.

%%%%%%%%%%%%%%%%%%%%%%%%%%%%%%%%%%%%%%%%%%%%%%%%%%%%%
%%%%%%%%%%%%%%%%%%%%%%%%%%%%%%%%%%%%%%%%%%%%%%%%%%%%%
\section{Mock data set}
\label{sec:sims}
%%%%%%%%%%%%%%%%%%%%%%%%%%%%%%%%%%%%%%%%%%%%%%%%%%%%%
%%%%%%%%%%%%%%%%%%%%%%%%%%%%%%%%%%%%%%%%%%%%%%%%%%%%%

Our study is performed on a simulated dataset of BBHM events obtained populating two of the {\sc eagle} simulations \citep{schaye2015} with merger events drawn from the \mobse{} population synthesis code \citep{Giacobbo:2017qhh} as described in~A19, A20. 

We mainly use the {\sc eagle} simulation with a box size of $100\,{\rm Mpc}$ and $1504^3$ gas and dark matter particles initially, in which the cosmology is fixed to the {\it Planck 2013} results~\citep{Planck:2013pxb}. We consider 7 redshift snapshots, specifically at $z\in \{0,\,1,\,2,\,3,\,4,\,5,\,6\}$. For each we use:
\begin{itemize}
    \item the full DM particle snapshot (namely, the position of all the DM particles in the box) to estimate the matter density field in section~\ref{sec:bias_mock};
    \item the host galaxy catalogue (describing the physical properties of the galaxies that host the mergers), which is a key ingredient for the semi-analytical model described in section~\ref{sec:bias_seman} and enters the implementation of the machine learning algorithm in section~\ref{sec:ML}.
\end{itemize}

Then we consider a BBHM catalogue as well, obtained by post processing the previously described snapshots with the population synthesis code \mobse{}. While \mobse{} alone provides binary catalogues only based on the population synthesis models, the post-processing procedure combines it with the {\sc eagle} stellar particles. This allows us to populate the galaxies inside the hydrodynamical simulation with BBHM (details in A19, A20); we refer to the output catalogues of this procedure as {\sc eagle}+\mobse{}. 
We use the BBHM data-set (which includes position and physical properties of all the merger events) to estimate the auto- and cross- power spectra of the tracer in section~\ref{sec:bias_mock}, and train and test our machine learning algorithm in section~\ref{sec:ML}. Note that in the former case we marginalize over all the properties of the mergers (e.g.,~progenitor masses, spins etc), while for the latter we include them explicitly; 

In the first part of our analysis, we develop a self-consistent algorithm to estimate the power spectrum and clustering properties (i.e.,~the bias) in a mock data-set. Due to the huge computational cost required to run the full hydrodynamics, the {\sc eagle} suite provides simulations with volumes of $(100\,{\rm Mpc})^3$, or smaller. This is limiting for our cosmological application, since the large linear scales are inaccessible from the simulation and non-linear effects become non-negligible, particularly at low redshift. As we discuss in section~\ref{sec:bias}, this makes it necessary to go beyond tree-level and jointly measure the linear bias $b_1$ together with higher order bias terms (see e.g., the review in~\citealt{Desjacques:2016bnm})
We validate the results of our algorithm by applying it to the study of DM halo clustering in the {\sc Quijote} simulation~\citep{villaescusa_2020}. {\sc Quijote} provides N-body DM-only snapshots at $z\in\{0,\,0.5,\,1,\,2,\,3\}$ inside a $L = 1000\,h^{-1}{\rm Mpc}$ comoving box. Using {\sc Quijote} we have access to large, linear scales, which allows us to directly extract $b_1$. In this way we can compare the results of our $b_1$ measurement at non-linear scales with tree-level power spectrum and standard bias estimators, checking the reliability of our pipeline. More detail on this validation test can be found in appendix~\ref{sec:validation}.

To calibrate the semi-analytical models in the second part of our analysis, we also use the {\sc eagle}+\mobse{} catalogues, in this case to extract the probability density functions (PDF) of galaxies and merger events, as a function of the stellar mass and star formation rate. 

Finally, to train and test the neural network that reproduces the map between the number of mergers and properties of their host, we use the aforementioned $L = 100\,{\rm Mpc}$ {\sc eagle}+\mobse{} data-set and a smaller realization based on the $25\,\rm Mpc$ side with initial $752^3$ gas and dark matter particles. 
%with the reference model~\citep{schaye2015}. 
In the following, we refer to them as EAGLE100 and EAGLE25, respectively.

%%%%%%%%%%%%%%%%%%%%%%%%%%%%%%%%%%%%%%%%%%%%%%%%%%%%%
%%%%%%%%%%%%%%%%%%%%%%%%%%%%%%%%%%%%%%%%%%%%%%%%%%%%%
\section{A short overview of cosmological bias}\label{sec:bias}
%%%%%%%%%%%%%%%%%%%%%%%%%%%%%%%%%%%%%%%%%%%%%%%%%%%%%
%%%%%%%%%%%%%%%%%%%%%%%%%%%%%%%%%%%%%%%%%%%%%%%%%%%%%

Forthcoming GW detectors will access almost the full sky, but they will be characterized by a low sky-localization precision (see e.g., the latest results by~\citealt{Iacovelli:2022bbs}). Therefore, a power spectrum analysis of GW clustering will be able to probe only large scale modes, which are in the linear regime. On such scales, for standard Gaussian initial conditions, the bias of a general DM tracer well described by a single constant coefficient, $b_1$. 
In this work, we want to directly measure $b_1$ for GW sources from cosmological, hydrodynamical simulations, to provide a benchmark for future studies. 

The simulation we analyse, however, is necessarily limited to small cosmological volumes and contains scales ranging from mildly to fully non-linear: the size of the simulated box sets $k_{\rm min} = 10^{-2}\,h{\rm Mpc}^{-1}$, while $k_{\rm max}$ is set by the Nyquist frequency, which in turn depends on how finely the mass distribution is sampled within the simulation. To avoid too large non-linearities, as we describe in the next section, we only take into account scales up to $\sim 0.3\,h{\rm Mpc}^{-1}$.
At these scales, perturbative bias models become significantly more complex and must include several higher-order parameters, that we have to measure and marginalize over, if we want to estimate $b_1$.\footnote{A more refined way to estimate $b_1$ from mock data, achieving higher precision and without including higher order coefficients, would be to rely on a separate Universe approach \citep{Mead:2013jta,Wagner:2014aka,Li:2015jsz} to extrapolate the currently available {\sc eagle}+\mobse{} results on larger scales. Unfortunately, the separate Universe methodology requires many simulations, generated for different input values of $\Omega_m$, which are not available for GW sources. In the present work we indeed rely on a single realization, at fixed fiducial cosmological parameters~\citep{Planck:2013pxb}.} 

In this work, we adopt the bias convention detailed in  \citet{Desjacques:2016bnm}, which collects earlier results from e.g. \citet{McDonald_2009,Chan:2012jj}. In this framework, the expansion of the halo density contrast $\delta_h$ up to third order can be written as
\begin{equation}\label{eq:bias_exp}
    \begin{split}
        \delta_h \sim\; & b_{1,h}\delta+b_{\nabla^2\delta,h}\nabla^2\delta+\epsilon\,+\\
        & +\frac{1}{2}b_{2,h}\delta^2+b_{K^2,h}(K_{ij})^2+\epsilon_\delta\delta\,+\\
        & +\frac{1}{6}b_{3,h}\delta^3+b_{\delta K^2,h}\delta(K_{ij})^2+b_{K^3,h}(K_{ij})^3+b_{\mathrm{td},h}O_\mathrm{td}^{(3)}\,+\\
        & +\epsilon_{\delta^2}\delta^2+\epsilon_{K^2}(K_{ij})^2\,,
    \end{split}
\end{equation}
where, for convenience, we dropped the time $\tau$ and scale $\boldsymbol{k}$ dependencies of the matter density field $\delta$ and of the bias operators and parameters. The first three lines captures correspondingly increasing orders in the bias expansion and contain local ($b_{i=1,2,...},\,b_{K^2}$),  non-local ($b_{\nabla^2\delta},\,b_\mathrm{td}$) and stochastic ($\epsilon_X$) contributions. 
This functional form actually  holds not just for halos but for a generic LSS tracer field $\delta_t$, irrespective of its nature; for this reason in the following we replace $h\to t$.

From Eq.~\eqref{eq:bias_exp}, the tracer auto-power spectrum, $P_{tt}(k)$, and its cross power spectrum with the matter field, $P_{t\text{DM}}(k)$, can be directly derived up to next-to-leading order in the expansion. These include terms up to fourth order in the $\delta$ perturbative expansion, and can be written as

\begin{equation}\label{eq:auto_exp}
    \begin{split}
        P_{tt}(k)=\; & b_{1,t}^2\big(P_L+2P_{13}+P_{22}\big)(k)\quad+ \\
        & +2b_{1,t}\sum_{O\in\{\delta^2,\,K^2,\,td\}}b_{O,t}(F_O+I_O)(k)\quad+ \\
        & +\sum_{O,\,O'\in\{\delta^2,\,K^2\}}b_{O,t}b_{O',t}I_{OO'}(k)\quad+\\
        & +P_\epsilon-2b_{1,t}b_{\nabla^2\delta,t}k^2P_L(k)\,,
    \end{split}
\end{equation}
\begin{equation}\label{eq:cross_exp}
    \begin{split}
    P_{t\text{DM}}(k)=\; & b_{1,t}\big(P_L+2P_{13}+P_{22}\big)(k)\quad+ \\
    & +\sum_{O\in\{\delta^2,\,K^2,\,td\}}b_{O,t}(F_O+I_O)(k)\quad- \\
    & -b_{\nabla^2\delta,t}k^2P_L(k)\,,
    \end{split}
\end{equation}
where $P_L$, $P_{13}$ and $P_{22}$ are the tree-level and 1-loop contributions to the matter power spectrum.
With respect to Eq.~\eqref{eq:bias_exp}, we here introduced $b_{\delta^2} =b_2/2$. 
All the other terms define the tree-level and 1-loop contributions to the matter power spectrum; we explicitly define them in appendix~\ref{sec:bias_ops} (for a complete derivation see \citet{Desjacques:2016bnm} and references therein).

We refer to the model in Eqs.~\eqref{eq:auto_exp} and~\eqref{eq:cross_exp} as {\sc EFTofLSS} and we use it as starting point for the mock-data bias analysis in section~\ref{sec:bias_mock}. We also consider a simpler model, in which third order and scale-dependent contributions are neglected. This is defined as

\begin{equation}\label{eq:auto_exp_limd}
    \begin{split}
        P_{tt}(k)=\; & b_{1,t}^2\big(P_L+2P_{13}+P_{22}\big)(k)\quad+ \\
        & +2b_{1,t}\sum_{O\in\{\delta^2,\,K^2\}}b_{O,t}(F_O+I_O)(k)\quad+ \\
        & +\sum_{O,\,O'\in\{\delta^2,\,K^2\}}b_{O,t}b_{O',t}I_{OO'}(k)+P_\epsilon\,,
    \end{split}
\end{equation}

\begin{equation}\label{eq:cross_exp_limd}
    \begin{split}
    P_{t\text{DM}}(k)=\; & b_{1,t}\big(P_L+2P_{13}+P_{22}\big)(k)\quad+ \\
    & +\sum_{O\in\{\delta^2,\,K^2\}}b_{O,t}(F_O+I_O)(k)\,.
    \end{split}
\end{equation}
This model is labelled {\sc LIMD} (Local In Matter Density), as it accounts only for local contributions of the matter field to the power spectrum.

In both Eqs.~\eqref{eq:auto_exp} and~\eqref{eq:auto_exp_limd}, the $P_\epsilon$ term represents the first order stochastic contribution.
This is scale independent and in first approximation corresponds to the shot noise $\bar{n}_t^{-1}$, where $\bar{n}_t$ represents the tracer density \citep{Dekel:1998eq}. To reduce the dimensionality of the problem, we neglect higher order, subdominat, stochastic terms by setting  $\epsilon_\delta=\epsilon_{\delta^2}=\epsilon_{K^2}=0$ and $P_\epsilon=\bar{n}_t^{-1}$. We also rewrite $b_{K^2}=-2(b_{1,t}-1)/7$ to further reduce the number free parameters in the analysis~\citep{baldauf_2012}. The interested reader can find further details on this part of the analysis in appendix~\ref{sec:estimators}.

Higher order correlators (e.g.,~the bispectrum~\citealt{Matarrese:1997sk}) are useful to measure $b_{1,t}$ on large scales (as it is done e.g.~in~\citealt{Jung:2022rtn}), and could be useful in this case as well to break degeneracies between bias parameters at different orders. However, because of the range of scales that we can safely access from the simulation ($10^{-2}\,h{\rm Mpc}^{-1}\lesssim k \lesssim 0.3\,h{\rm Mpc}^{-1}$), they should be expanded beyond the tree-level as well. This would make their implementation particularly challenging; moreover, their application to real data analysis would be limited by the lack of sky-localization in GW detection. For these reason, we leave further investigation on this point for a follow-up work.

%%%%%%%%%%%%%%%%%%%%%%%%%%%%%%%%%%%%%%%%%%%%%%%%%%%%%
%%%%%%%%%%%%%%%%%%%%%%%%%%%%%%%%%%%%%%%%%%%%%%%%%%%%%
\section{Methodology}\label{sec:method}
%%%%%%%%%%%%%%%%%%%%%%%%%%%%%%%%%%%%%%%%%%%%%%%%%%%%%
%%%%%%%%%%%%%%%%%%%%%%%%%%%%%%%%%%%%%%%%%%%%%%%%%%%%%

The linear bias parameter of binary black hole mergers can be estimated from the data-sets presented in section \ref{sec:sims} by determining the tracer power spectra from the catalogues and fitting them with the models discussed in section \ref{sec:bias}. We present our $b_{1,m}$  estimates in section \ref{sec:bias_mock} and we test our results against semi-analytical models in section \ref{sec:bias_seman}, as a counter-check of the reliability of this approach. 

In this way, we are also able to understand which physical parameters are crucial to determine the clustering of the sources. Such analysis can then be compared with the outputs of the machine learning algorithm described in section \ref{sec:ML}, in which the relative importance of the parameters is tested inside the simulation itself.

\subsection{Direct estimation on mock data}\label{sec:bias_mock}

To measure the bias of any given tracer $t$ in our mock data-sets, we first extract the tracer power spectrum $P(k)$, by relying on a straightforward estimator that can be summarized as follows: 

\begin{enumerate}
    \item We estimate the number density field $\tilde{n}(\boldsymbol{r})$ by %in the snapshot 
    weighing the particle positions with a Cloud-In-Cell (CIC) %weighting 
    scheme.
    \item We compute the density contrast $\tilde{\delta}(\boldsymbol{r})=\tilde{n}(\boldsymbol{r})/\bar{\tilde{n}}-1$ ($\bar{.}$ denotes spatial average) and we perform a fast Fourier transform to get $\tilde{\delta}(\boldsymbol{k})$.
    \item We deconvolve the CIC filter function from the estimated $\tilde{\delta}(\boldsymbol{k})$ to recover the actual Fourier transform of the field, $\delta(\boldsymbol{k})$. 
    \item We measure the auto- or cross- tracer-DM power spectrum estimator, $\tilde{P}_{xy}$, by averaging $\delta_{x}(\boldsymbol{k})\delta_{y}^*(\boldsymbol{k})$, where $x,y$ can be either dark matter(DM)  or the tracer ($t$), inside bins in Fourier space and estimate the $1\sigma$ error bars on $\tilde{P}_{xy}(k)$ by taking the square root of the Gaussian variance of our results.
\end{enumerate}
As a counter-check, we also implement a second estimator, based on the FKP prescription ~\citet{Feldman:1993ky}. Further details on the technical implementation of the two estimators are discussed in appendix~\ref{sec:estimators}. 

Having obtained estimates $\hat{P}_{tt}(k)$ and $\hat{P}_{t\text{DM}}(k)$, we then fit the models of Eqs.~\eqref{eq:auto_exp}-\eqref{eq:cross_exp_limd} to our results via Markov Chain Monte Carlo. We assume a Gaussian likelihood for $\hat{P}(k)$
and consider the parameter vectors $(b_{1,t},\,b_{2,t},\,b_{\mathrm{td},t},\,b_{\nabla^2\delta,t})$ and $\boldsymbol{\theta}=(b_{1,t},\,b_{2,t})$, for the {\sc EFTofLSS} and {\sc LIMD} models, respectively;
we then take a uniform prior in the interval $b_{1,t}\in[0,\,10]$ for the linear bias parameter and uniform improper priors on all other parameters. 
We combine the auto- spectra corrected for the shot noise (that is, we subtract $P_\epsilon=\bar{n}_t^{-1}$) with the cross-spectra and truncate the analysis at a given scale, $k_\mathrm{NL}$ (specified later), chosen to exclude the highly non-linear regime where perturbation theory fails. Posterior sampling is performed with the \texttt{emcee}\footnote{\url{https://github.com/dfm/emcee}.} package \citep{foreman_2013}, while for the computation of the bias operators $I_O$, $F_O$ and $I_{OO'}$ that appear in the {\sc EFTofLSS} and {\sc LIMD} models (see appendix~\ref{sec:bias_ops}) we use \texttt{velocileptors}\footnote{\url{https://github.com/sfschen/velocileptors}.} \citep{Chen:2020fxs,Chen:2020zjt}.
To constrain $b_{1,t}$, we treat all the non-linear bias parameters as nuisance parameters and marginalize over them. Our final results for are maximum a posteriori estimates and we use the 18th and 84th percentiles to define our $b_{1,t}$ credible interval (see Figs.~\ref{fig:corner_bbh_z0} and~\ref{fig:corner_bbh_z6}).
The reliability of our bias estimator is preliminarly tested on DM halo catalogues from the {\sc Quijote} simulations, for which we can easily compare our results with theoretical expectations (see appendix~\ref{sec:validation}).

\begin{table}
\centering
\renewcommand{\arraystretch}{1.2}
\setlength{\tabcolsep}{10pt}
\begin{tabular}{l|c|c}
$z$ & \textsc{EFTofLSS} $b_{1,m}$ & \textsc{LIMD} $b_{1,m}$ \\
\hline
0 & $1.00_{-0.44}^{+1.92}$ & $1.29_{-0.84}^{+0.09}$ \\[5pt]
1 & $1.21_{-0.54}^{+1.78}$ & $1.81_{-0.93}^{+0.07}$ \\[5pt]
2 & $1.35_{-0.56}^{+2.36}$ & $2.23_{-0.88}^{+0.13}$ \\[5pt]
3 & $1.73_{-0.65}^{+2.80}$ & $2.67_{-0.91}^{+0.22}$ \\[5pt]
4 & $2.35_{-0.65}^{+3.11}$ & $3.11_{-0.96}^{+0.24}$ \\[5pt]
5 & $4.50_{-2.14}^{+1.76}$ & $3.49_{-0.99}^{+0.33}$ \\[5pt]
6 & $5.73_{-2.81}^{+1.49}$ & $3.80_{-0.89}^{+0.56}$
\end{tabular}
\caption{$b_{1,m}$ estimates obtained via MCMC on {\sc eagle}+\mobse{} data-sets.}
\label{tab:MCMC_b1}
\end{table}

We then apply our algorithm to estimate the linear bias of BBHM at redshifts $z\in\{0,\,1,\,2,\,3,\,4,\,5,\,6\}$. In this analysis, we truncate the BBHM auto- and BBHM-DM cross-spectra at $k_\mathrm{NL}=0.25\,h\,$Mpc$^{-1}$ and $k_\mathrm{NL}=0.35\,h\,$Mpc$^{-1}$, for $z=0$ and $z=\{1-6\}$, respectively. 
The results of the analysis are summarized in Table \ref{tab:MCMC_b1} and Fig.~\ref{fig:merger_bias_vs_HOD}, which respectively show and plot the values of $b_{1,m}$ and the corresponding errors in the chosen redshift bins.

%%%%%%%%%%%%%%%%%%%%%%%%%%%%%%%%%%%%%%%%%%%%%%%%%%%%%
%%%%%%%%%%%%%%%%%%%%%%%%%%%%%%%%%%%%%%%%%%%%%%%%%%%%%
\subsection{Semi-analytical model}
\label{sec:bias_seman}
%%%%%%%%%%%%%%%%%%%%%%%%%%%%%%%%%%%%%%%%%%%%%%%%%%%%%
%%%%%%%%%%%%%%%%%%%%%%%%%%%%%%%%%%%%%%%%%%%%%%%%%%%%%

To validate the results of the previous section, we also evaluate the bias with a different approach, which exploits information extracted from the simulations as an input to build a Halo Occupation Distribution model (HOD, see e.g.,~\citealt{Peacock:2000qk,Berlind:2001xk,Karagiannis:2018jdt,Libanore:2020fim}) and connect the BBHM bias with the bias of their host galaxies and DM halos. 

In~\citet{Libanore:2020fim}, a similar technique was firstly applied to the study of merger bias. With respect to the usual HOD approach~\citep{Peacock:2000qk,Berlind:2001xk} --- aimed at modeling the bias of galaxies from the distribution of their host DM halos -- this now requires an extra step to account for the binary black hole distribution as a function of the host galaxy features; all the relevant distributions are calibrated using our mock \textsc{eagle}+\mobse{} dataset, also analysed in section~\ref{sec:bias_mock}. Compared to~\citet{Libanore:2020fim}, we now adopt 
an updated simulation setup, using the $L = 100$\,Mpc side {\sc eagle} box (while~\citealt{Libanore:2020fim} used data from the 25\,Mpc box). We also use the full mock data-set (the same as in section~\ref{sec:bias_mock}), while in~\citet{Libanore:2020fim} specific selection effects due to the Einstein Telescope signal-to-noise ratio were implemented.

We now discuss the aforementioned procedure more in detail. First of all, we need to compute the bias of the host galaxies, which requires the following three ingredients: 
\begin{enumerate}
    \item the halo mass function $dn(z)/dM_h$ in each snapshot, i.e.,~the number of halos per mass bin. We consider 6 halo mass bins with $M_h$ uniformly spaced in log-space between $10^{9}\,M_\odot$ and $4\times 10^{14}\,M_\odot$. 
    \item the linear halo bias $b_{1,h}(M_h,\,z)$, for which we rely on the parametrization derived in~\citep{Tinker:2010my} using the peak-background split approach. In appendix~\ref{sec:validation} --Figs.~\ref{fig:quijote_halo_bias_LIMD} and~\ref{fig:quijote_halo_bias_EFTofLSS} --
    we compare this parametrization with the actual estimates of $b_{1,h}$ from simulations at $z =\{0,\,3\}$, obtained from the algorithm described in the previous section. 
    \item the galaxy HOD $\langle N_g(M_*,\psi,z)|M_h\rangle$, namely the probability of finding $N_g$ galaxies having stellar mass $M_*$ and star formation rate $\psi$ inside a halo of mass $M_h$ in the $z$ snapshot. The number per bin is directly estimated from the simulations; we consider 15 stellar mass bins between $ M_* = 10^{7}\,M_\odot$ and $10^{12}\,M_\odot$ and 15 star formation rate bins between $\psi = 10^{-5.5}\,M_\odot{\rm yr}^{-1}$ and $10^{2.5}\,M_\odot{\rm yr}^{-1}$.
\end{enumerate}
We note that in~\citet{Libanore:2020fim} a similar procedure was applied by fitting data with the \citet{Tinker:2010my} halo mass function and the {\sc eagle} HOD analytical prescription~\citep{10.1093/mnras/sty2110}. 
We verified that such prescription holds as a first-order approximation, but using the halo and galaxy PDFs directly measured on simulations is more reliable when considering bins with a small number of data. 

By combining all these ingredients, we compute the galaxy linear bias per stellar mass and star formation rate bin at each redshift as 
\begin{equation}\label{eq:bias_gal}
    b_{1,g}(M_*,\psi,z) = \frac{\int dM_h\,b_{1,h}(M_h,z)\frac{dn(z)}{dM_h}\langle N_g(M_*,\psi,z)|M_h\rangle}{\int dM_h\,\frac{dn(z)}{dM_h}\langle N_g(M_*,\psi,z)|M_h\rangle},
\end{equation}
where the numerical integration is performed between the values of $M_h^{min}$ and $M_h^{max}$ that define the halo mass bins.

As mentioned, the semi-analytical model we build to describe the merger bias adds an extra layer to the analysis discussed up to this point, namely the galaxy occupation distribution $\langle N_m(z) | M_*,\psi \rangle$, which we define as the probability of finding $N_m$ mergers in a galaxy as a function of its stellar mass $M_*$ and star formation rate $\psi$. We estimate $\langle N_m(z) | M_*,\psi \rangle$ from the {\sc eagle}+\mobse{} simulation and finally model the linear bias of the merger as 
\begin{equation}\label{eq:merger_bias}
    b_{1,m}(z) = \frac{\int dM_*\int d\psi\,b_{1,g}(M_*,\,\psi,\,z)\frac{dn_g(z)}{dM_*d\psi}\,\langle N_m(z) | M_*,\psi \rangle}{\int dM_*\int d\psi\,\frac{dn_g(z)}{dM_*d\psi}\,\langle N_m(z) | M_*,\psi \rangle} \,,
\end{equation}
and also in this case the numerical integration is performed between the $M_*$ and $\psi$ bins boundaries. The galaxy number density $dn_g(z)/dM_*d\psi$ is computed by dividing the number of galaxies in each bin $N_g(M_*,\psi,z)$ by the width of the $\Delta M_*\Delta\psi$ bin and the volume of the simulation box, $L^3$.
We checked that marginalizing over the star formation rate dependence before computing the galaxy and merger bias (i.e.,~dividing the galaxies and merger only in dependence of $M_h,\,M_*$) has a small impact on the final results. Indeed, as we discuss in detail in section~\ref{sec:ML}, the merger distribution properties mainly depend on the stellar mass (this is in agreement to what was initially pointed out in~\citet{artale_10.1093/mnras/stz1382}, A19 and then confirmed in~\citealt{Santoliquido:2022kyu}).
The HOD contains an intrinsic uncertainty due to the theoretical assumptions in the model (e.g.,~assuming dependence only on the halo mass, stellar mass and star formation rate). With this caveat in mind, we verified that the statistical error generated by propagating\footnote{The error estimate was computed by assuming that $N_{g}$ and $N_m$ are independent, Gaussian random variables. We apply the standard error propagation formulae separately on the numerator and denominator, assuming that the contribution from $dn/dM_h$ and $b_{1,h}(M_h,z)$ is negligible. The ratio between these two quantities is described by a Cauchy distribution, for which the expected value and variance are undefined; we overcome this problem relying on the prescriptions in~\citet{JSSv016i04}. } the uncertainties on $\langle N_g(M_*,\psi,z) | M_h \rangle$ and $\langle N_m(z) | M_*,\psi \rangle$ in eqs.~\eqref{eq:bias_gal} and~\eqref{eq:merger_bias} is negligible compared to the errorbars provided by the numerical estimate of $b_{1,\,m}$ in the previous section.
The internal comparison between the semi-analytical linear bias of the mergers estimated from Eq.~\eqref{eq:merger_bias} and the results from section~\ref{sec:bias_mock} is displayed in Fig.~\ref{fig:merger_bias_vs_HOD}.

%%%%%%%%%%%%%%%%%%%%%%%%%%%%%%%%%%%%%%%%%%%%%%%%%%%%%
%%%%%%%%%%%%%%%%%%%%%%%%%%%%%%%%%%%%%%%%%%%%%%%%%%%%%
\subsection{Machine learning methodology}
\label{sec:ML}
%%%%%%%%%%%%%%%%%%%%%%%%%%%%%%%%%%%%%%%%%%%%%%%%%%%%%
%%%%%%%%%%%%%%%%%%%%%%%%%%%%%%%%%%%%%%%%%%%%%%%%%%%%%

The numerical simulations allow us to verify which physical parameters are the most relevant in the modelling of BBHM distribution and their spatial clustering. The former was previously studied in A19, which our results are in agreement with.
To do so, we train a neural network to paint BBHM events on the simulations on a galaxy by galaxy basis depending on the physical properties of their host, in a similar fashion to how HOD are routinely used, following the approach established since \citet{Kamdar:2015fla}.
For example, knowing the characteristics of a certain galaxy (e.g.,~stellar mass, star formation rate, metallicity...), the NN determines how many BBHM it is likely to contain. In this way, our NN can be used to study the host parameters relative importance for the presence of BBHM, as well as can be applied to quickly paint them on N-body or  cosmological, hydrodynamical simulations.
It is known \citep{Kamdar:2015fla, Agarwal:2017xas, deSanti:2022jlq} how this procedure tends to successfully reproduce the average properties of the target parameters, but fails in reproducing their distributions. In our case, the BBHM have an intrinsic scatter due to both the probabilistic nature of the events, and the baryonic physics that is not easily captured by galaxy properties~\citep{Stiskalek:2022nsr}.
Therefore, we will compare the performance of a NN trained to assign to each galaxy a number of BBHM in a deterministic way, with the performance of a NN trained to assign to each galaxy the BBHM PDF, as done in \citet{374138, Villaescusa-Navarro:2021pkb, Stiskalek:2022nsr}, based on their ability to reproduce the BBHM global PDF and power spectrum. In the following we will dub the two methods \emph{deterministic} and \emph{statistical}, respectively.
To account for the strong dependence of the BBHM number on the galaxy mass $M_\star$,\footnote{We define the galaxy mass as the sum of the bounded stellar particle masses. The same argument applies to any other definition of galaxy mass.}
we fit the pairs ($M_\star$, BBHM number) of the training set with a cubic polynomial and calculate the fractional residuals dividing the BBHM number data by the interpolant, in order to remove the main scaling. The NN is then trained on the fractional residuals. 
This is not only motivated by speeding up the NN training, but also from the fact that the prediction from the simple cubic interpolation already leads to a reasonable approximation of the power spectrum (see section~ \ref{sec:results}).

We randomly divide the catalogues from EAGLE100 into training (60\%), validation (20\%) and test (20\%) sets. This leads to a spatially inconsistent distribution of the galaxies of each set; while we can assume that each set is a fair sample of the BBHM PDF, none of them can be used by itself to calculate a meaningful power spectrum. Thus, the EAGLE25 is used to test the power spectrum reconstruction instead.
We test different parametrizations of the BBHM PDF and the PDF of their fractional residuals, finding that a Gaussian approximation of the fractional residuals offers the best results.\footnote{We use the normal, lognormal and Poisson distributions for the BBHM PDF, while we do not apply the Poisson distribution to the residuals since this is not meaningful for PDF obtained as the ratio of two number of events.}

Both the deterministic NN and the statistical NN share the same architecture for the input and hidden layers, while the output layers differs. In the former case we have a single node output with \texttt{elu}+1 activation both when we are fitting for the BBHM number and the fractional residuals. In the case of normal and lognormal distributions, we use a two node output: the first is interpreted as mean and the second as standard deviation of the variable (or logarithm of the variable, respectively). In the case of Poisson distribution, the single output is the mean of the distribution. In all output nodes an \texttt{elu}+1 activation function is used again, with the excepton of the mean of the lognormal distribution which has a linear activation.
For the deterministic network, we use the mean square errors as loss function, while in the statistical network the likelihood of the parametrized distribution is used instead.
As optimizer, we employ \texttt{Adam} \citep{2014arXiv1412.6980K} with a cyclical learning rate \citep{2015arXiv150601186S}, and we apply a dropout \citep{JMLR:v15:srivastava14a} to all hidden layers.
We tune the number and hidden layers, nodes, dropout rate and activation functions, setting for 3 hidden layers of 128 nodes, a dropout rate of 0.1 and \texttt{selu} \citep{2017arXiv170602515K} activation functions. 

\begin{table*}
    \centering
    \begin{tabular}{|lll|}
    \hline
              & Merger tree info & Description \\
    \hline
        Host galaxy parameter & & \\
    \hline
        MassType\_Star, $M_\star$ [$M_\odot$]   & \cmark & Total stellar mass\\
        SF\_Metallicity, $Z$                    & \cmark & Metal mass fraction in the star-forming gas \\
        StarFormationRate, $\psi$ [$M_\odot$/yr] & \cmark & Total star formation rate \\
        Vmax, $V_\text{max}$ [km/s]                  & \cmark & Maximum of the circular velocity rotation curve \\
        StellarVelDisp  [km/s]                  & \xmark & Average one dimensional star velocity dispersion \\
        TotalEnergy     [$M_\odot$ km$^2$/s$^2$]& \xmark & Total energy of the galaxy \\
        KineticEnergy                           & \xmark & Total kinetic energy \\
    \hline
        Host halo parameter & & \\
    \hline
        GroupMass, $M_H$ [$M_\odot$]            & \cmark & Total Firends-of-Friends mass\\
        Group\_M\_Crit200  [$M_\odot$]          & \cmark & Total mass within the radius containing a density 200 times the critical density of the Universe \\
        Group\_M\_Crit500  [$M_\odot$]          & \cmark & Total mass within the radius containing a density 500 times the critical density of the Universe \\
        Group\_M\_Crit2500  [$M_\odot$]         & \cmark & Total mass within the radius containing a density 2500 times the critical density of the Universe \\
    \hline
    \end{tabular}
    \caption{Parameters in the {\sc eagle} catalogue used as input features for the NN~\citep{McAlpine_2016}.
    If the parameter is used in the text, the string used in the {\sc eagle} database is paired with the symbol we use.
    The merger tree info are the value of said parameters for the object progenitors (if present) along the main branch within the 16 snapshots preceding the merger.
    }
    \label{tab:NN_feature_list}
\end{table*}

Since our goal is to verify how all the information on the BBHM distribution is encoded in a few parameters, we use a rather agnostic approach in choosing our baseline feature set. This includes the comprehensive list of halo and galaxy parameters available in the {\sc eagle} simulation, shown in Tab.~\ref{tab:NN_feature_list}, some of which we expect to have negligible effect on the BBHM distribution.

%%%%%%%%%%%%%%%%%%%%%%%%%%%%%%%%%%%%%%%%%%%%%%%%%%%%%
%%%%%%%%%%%%%%%%%%%%%%%%%%%%%%%%%%%%%%%%%%%%%%%%%%%%%
\section{Results}\label{sec:results}
%%%%%%%%%%%%%%%%%%%%%%%%%%%%%%%%%%%%%%%%%%%%%%%%%%%%%
%%%%%%%%%%%%%%%%%%%%%%%%%%%%%%%%%%%%%%%%%%%%%%%%%%%%%

\subsection{Bias estimates}

\begin{figure}
    \centering
    \includegraphics[width=\columnwidth]{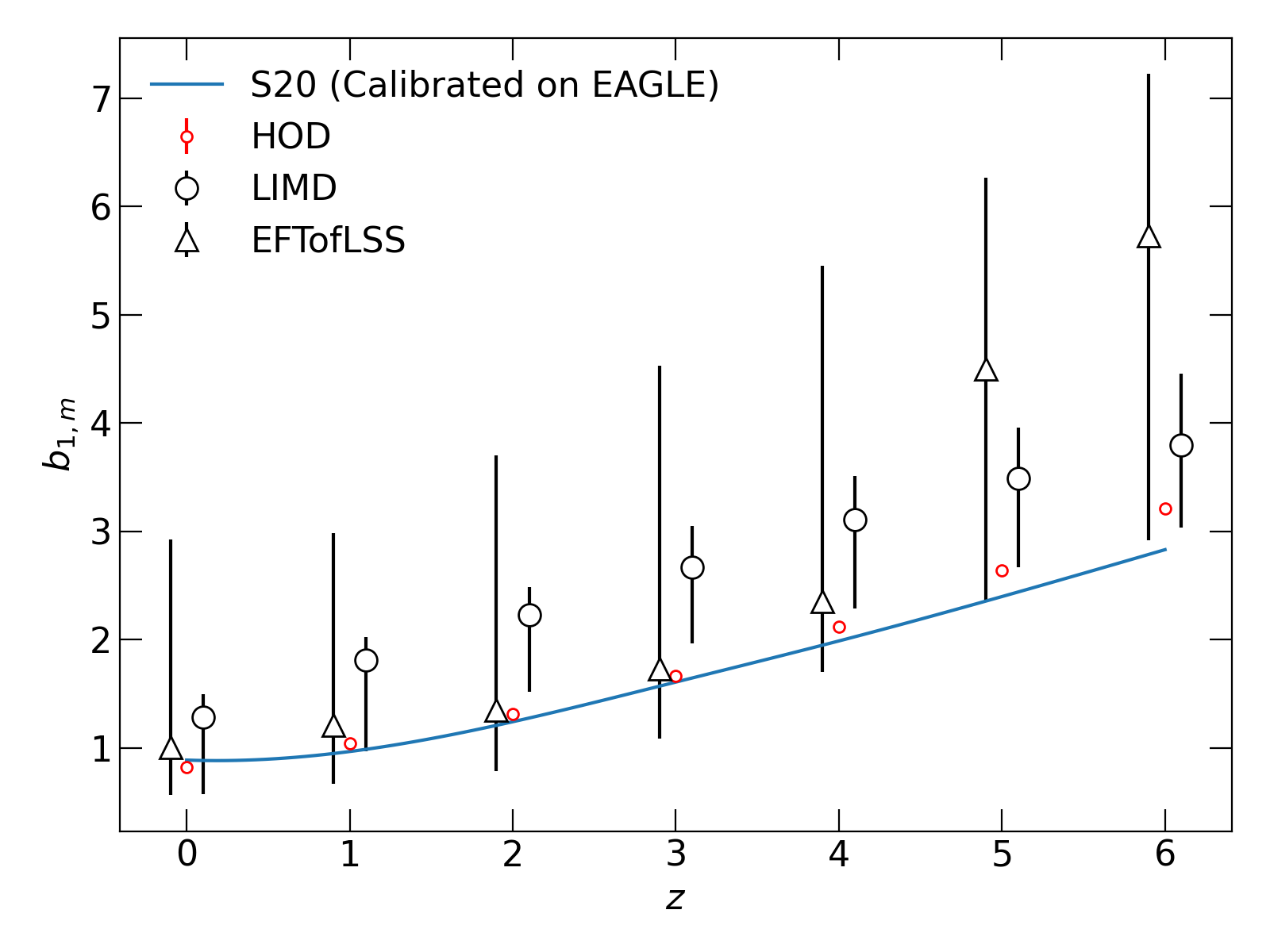}
    \caption{Bias estimates computed with the different methods described in the text: MCMC results from section~\ref{sec:bias_mock} and Tab.~\ref{tab:MCMC_b1} in the {\sc LIMD} (black circles) and {\sc EFTofLSS} (black triangles) cases, semi-analitycal HOD model from section~\ref{sec:bias_seman} (red circles) and updated merger rate-weighted results from section~\ref{sec:bias_rate} (S20, blue line).}
    \label{fig:merger_bias_vs_HOD}
\end{figure}

In section~\ref{sec:bias_mock} and~\ref{sec:bias_seman}, we have presented two independent tools to estimate the bias of BBHM from simulations, the former fully based on mock data-sets, while the latter requiring some analytical prescription to connect the mergers to the properties of their host galaxies. Tab.~\ref{tab:MCMC_b1} collects our results on the direct estimates, while Fig.~\ref{fig:merger_bias_vs_HOD} compares them with the semi-analytical results (whose error bars are negligible with respect to the numerical model ones, as we described in section~\ref{sec:bias_seman}). Results are in good agreement between the two methods and with the close-to-linear trend previously obtained by~\citet{Libanore:2020fim,Libanore:2021jqv}.  

For practical purposes and applications in follow-up analysis, we provide the following fit to the bias estimated from simulations:
\begin{equation}\label{eq:fit}
    b_{1,m} = 1.20 \, (1+z)^{0.59}\,.
\end{equation}
The parameters in the previous equation are calibrated on the {\sc LIMD} results, since we find that the {\sc LIMD} model is favored by the Bayes Information Criterion over the {\sc EFTofLSS} one.

\begin{figure}
    \centering
    \includegraphics[width=\columnwidth]{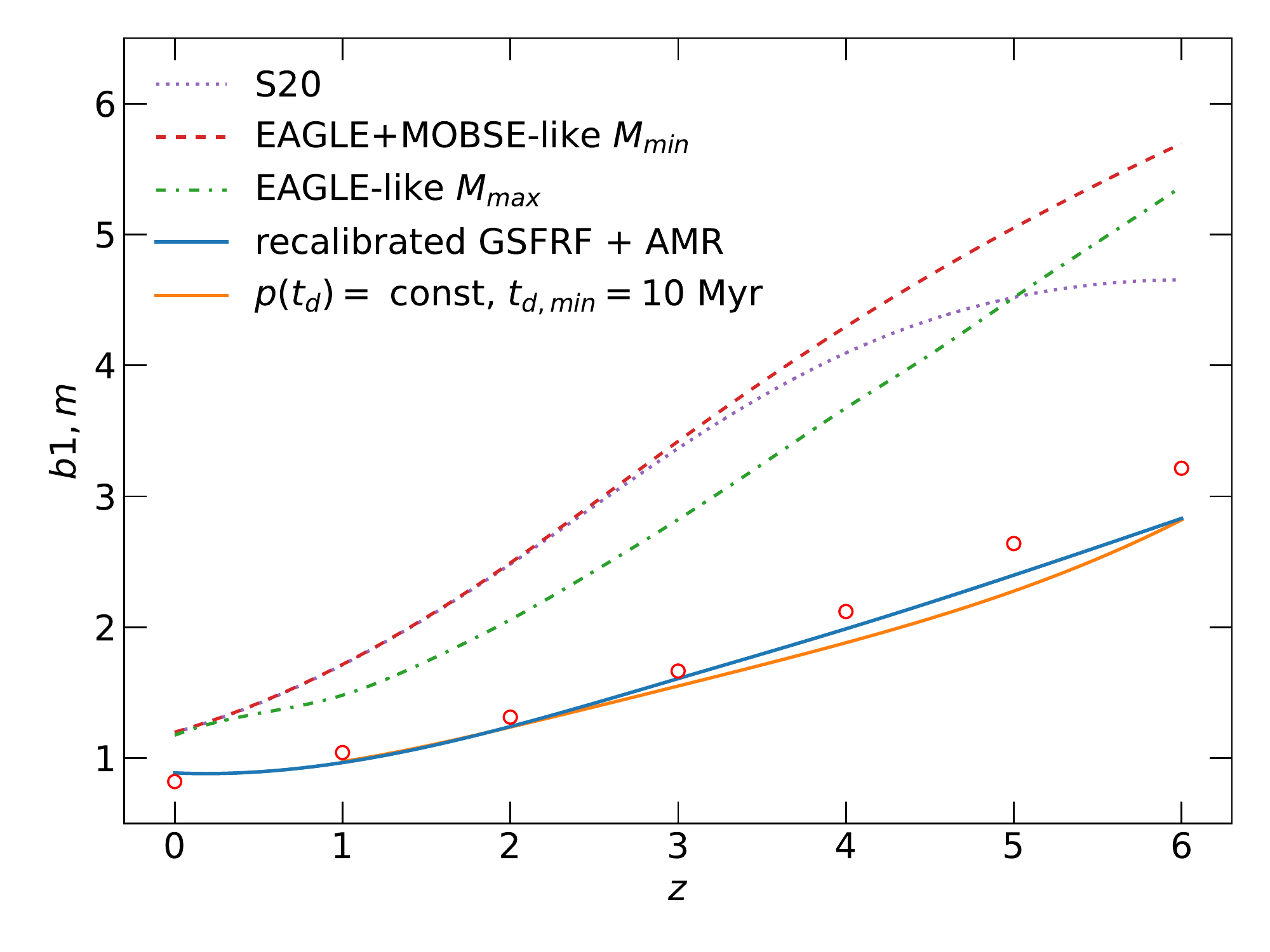}
    \caption{HOD bias values from Eq.~\eqref{eq:merger_bias} (red circles) compared with the merger rate-weighted biases computed via Eq.~\eqref{eq:merger_bias_scelfo} under different prescriptions.
    The blue thick line uses the same setup as S20, while dashed lines change the lower (orange) and upper (green) halo mass integration limits. We obtained the dotted red line by re-calibrating the star formation rate and abundance matching results on the {\sc eagle} catalogues; the purple dotted line relies on a different time delay distribution. More detail in the text.}
    \label{fig:S20_comparison}
\end{figure}

%%%%%%%%%%%%%%%%%%%%%%%%%%%%%%%%%%%%%%%%%%%%%%%%%%%%%
\subsection{Comparison with the merger-rate-weighted bias}
\label{sec:bias_rate}
%%%%%%%%%%%%%%%%%%%%%%%%%%%%%%%%%%%%%%%%%%%%%%%%%%%%%

The main goal of our analysis is to understand how much the BBHM bias is affected by the properties of the hosts. This is a key information to understand whether the clustering can be used to disentangle different host populations (and possibly formation channels), as well as to check how much our results depend on the specific astrophysical models assumed in the simulations. 
With this in mind and to verify the robustness of our results we re-estimated the bias following a completely independent approach with respect to the one adopted in previous section. This alternative methodology is based on the analysis depicted in~\citet{Boco_2019,Scelfo:2020jyw} (a similar approach is also followed by~\citet{Bellomo:2021mer}), where the BBHM bias is computed by weighting the bias of the hosts through the merger rate, which in turn relates with the host properties, modelled from theoretical assumptions and real data. 

As we will discuss in detail at the end of this section, the parametrization used in~\citet{Boco_2019,Scelfo:2020jyw} (hereafter B19, S20) is different with respect to the assumptions made by the authors of A19 to simulate the BBHM catalogues we rely on. This propagates to the bias computation, leading to %some
discrepancies between the results %obtained following 
of the two procedures (particularly at high redshift). Indeed, the bias in S20, as well as the bias in~\citet{scelfo2018}, 
is at face value 
consistent neither with each other, nor with our results.

To shed light on this possible issue, we focus on S20 and investigate the reason for the discrepancy.
We model the bias according to the procedure described in S20, to which we refer the reader for details, while we 
limit ourselves here to briefly recap the necessary building blocks.
In the same spirit as Eqs.~\eqref{eq:bias_gal},~\eqref{eq:merger_bias}, the bias is calculated convolving the halo mass function and bias with an appropriate kernel, describing the number distribution of BBHM, conditioned over the halo masses. 

We model the chirp mass distribution of the merging black holes %$dp/d\Mchirp$
as a function of the mass of the progenitor star and the metallicity of the host galaxy, as it is done in B19.
We adopt the initial mass function from~\citet{Chabrier:2003vv} %for $m_\star$ 
and we relate it to the black hole mass that the star forms as in~\citet{Spera:2017fyx}.
The galaxy distribution is modelled according to the galaxy star formation rate function (GSFRF) parametrized as a Schechter function in \cite{Mancuso_2016}.

All the quantities in the analysis are estimated at the binary formation time: this requires us to introduce a reasonable distribution for the time delay $t_d$ between this moment and the time of the merger.
The isolated binary evolution predicts the time delay distribution $p(t_d) \propto t_d^{-1}$, with a minimum time delay cutoff of $t_{d,\text{min}} = \SI{50}{Myr}$, which is used in n \{B19, S20\}.
In general, the exact form depends on several physical processes such as mass transfer, kicks, stellar winds, and orbital separation, and early type galaxies present a flatter distribution compared to early type galaxies \citep{Santoliquido:2022kyu}.
In our analysis, therefore, we test two independent cases: $p(t_d)\propto t_d^{-1}$ (to be consistent with S20), and a uniform distribution $p(t_d) \equiv \text{const}$ that recalls 
the one implemented in the simulations by A19 analysed in section~\ref{sec:bias_mock}. 
The specific choice of $p(t_d)$ may largely affect the final results on the black hole merger rate~\citep{2021MNRAS.502.4877S}, but how this propagates to the clustering of events, has not been discussed yet. We anticipate that our analysis shows that different choices of $p(t_d)$ have a limited impact on the bias (see Fig.~\ref{fig:S20_comparison}).

In this bias model, the host halo and host galaxy mass are related to the star formation rate via the abundance matching relations (AMR) of \cite{Aversa_2015}.

Having said that, we estimate the black hole merger rate $\mathcal{R}_{BH}(z,\log_{10}\psi,\,\Mchirp)$
as in \{B19, S20\} by marginalizing over the chirp mass $\Mchirp$ of the events and the star formation rate $\psi$ of the host galaxies. 
Finally, we estimate the merger bias through
\begin{equation}\label{eq:merger_bias_scelfo}
    b_{\rm 1,m}(z)
    =
    \frac{
        \int \diff \log_{10} \psi\,
        b_{1,g}(z,\psi)
        \int \diff \Mchirp\,
        \psi\,\mathcal{R}_{BH}(z,\log_{10}\psi,\Mchirp)}{
        \int \diff \log_{10} \psi
        \int \diff \Mchirp\,
        \psi\,\mathcal{R}_{BH}(z,\log_{10}\psi,\Mchirp)
    } \,,
\end{equation}
where the galaxy bias as a function of $\psi$ is compute as in~\citet{Aversa_2015},
i.e.,~$b_g(\psi,z) = b_{1,h}(M_h,z)$, where the halo bias is the same we discussed in the previous section. 

Different prescriptions on the models and parameters in Eq.~\eqref{eq:merger_bias_scelfo} can lead to very different bias results. We highlight this is Fig.~\ref{fig:S20_comparison}, in which the blue thick line has been recomputed using the same prescriptions as S20, the red circles are the bias values estimated in section~\ref{sec:bias_seman}, and we obtained all the other lines by tuning Eq.~\eqref{eq:merger_bias_scelfo}.

First of all, results in S20 include detector selection effects via a signal-to-noise ratio distribution; to relate to our analysis (which involves the intrinsic signal of the full merger population), we need to drop this term.
Then, in S20 the lower bound of the star formation rate integral is directly related to a lower bound on the mass of the host galaxy via the abundance matching relation. At every redshift this lower bound is chosen such that the integral converges.
Instead, in {\sc eagle}+\mobse{} at every redshift only galaxies with stellar mass $\gtrsim 10^7 \, M_\odot$ are included because of mass resolution limits (see,~e.g.,~the discussion in~\citealt{Mapelli:2018okv}) 
If the same cut is imposed in S20 (orange dashed line in Fig.~\ref{fig:S20_comparison}) the characteristic flattening S20 had at high redshift partially disappears, since the least biased objects are removed from the distribution.
The opposite effect, i.e. a decrease in the bias due to the absence of the most biased objects, takes place when we remove host galaxies with mass $\gtrsim 10^{12} M_\odot$. This is not due to a physically motivated effect, but to the fact that the {\sc eagle} simulation is quite limited in size, and does not contain larger objects. To circumvent this limitation, larger cosmological, hydrodynamical simulations or self-consistent extrapolation to higher massess are required.
Finally, we have to take into account that both the galaxy star formation rate function of \citet{Mancuso_2016} and the abundance matching relations provided in \citet{Aversa_2015} do not fully agree with the corresponding quantities extracted from {\sc eagle}. Re-calibrating these functions 
on the {\sc eagle} catalogues,
the bias estimated from Eq.~\eqref{eq:merger_bias_scelfo} turns out to be in broad agreement with our HOD results from Eq.~\eqref{eq:merger_bias} (red dotted line in Fig.~\ref{fig:S20_comparison}).
As anticipated, both to test how strong the dependence of the bias on the time delay distribution is, and to check consistency with the {\sc eagle}+\mobse{} simulation, we repeat the calculation using a different $p(t_d)$.
We pick a flat distribution with a minimum time delay of $t_{d, \text{min}} = \SI{10}{Myr}$ from the distributions considered in \cite{2021MNRAS.502.4877S}, obtaining the purple dotted line in Fig.~\ref{fig:S20_comparison}.
The two bias distributions ($p(t_d)\propto t_d$, and $p(t_d) \equiv $const) braket the one used in {\sc eagle}+\mobse{}. The difference between the two resulting bias is negligible; therefore, as we anticipated, we conclude that the BBHM clustering is only marginally affected by the time delay specific choice.

%%%%%%%%%%%%%%%%%%%%%%%%%%%%%%%%%%%%%%%%%%%%%%%%%%%%%
\subsection{Validation of the NN and its architecture}\label{sec:NN}
%%%%%%%%%%%%%%%%%%%%%%%%%%%%%%%%%%%%%%%%%%%%%%%%%%%%%

We start by discussing the necessity to move past the description of the average quantity in the context of ML-based painting of simulations, as also pointed out in \citet{Stiskalek:2022nsr}. As anticipated, a limited amount of physical properties of the host halo and, in our case, of the host galaxy, fail to describe \emph{the} single realization of the painted label. Instead, what they characterize is the label probability density function: this is not at all dissimilar to the standard HOD approach.
Employing the NN to learn the PDF, rather than the single realization of the label, not only allows us to better reconstruct its variance, %of the label ---admittedly a rather tautological finding--- 
which is often underestimated~\citep{Kamdar:2015fla, Agarwal:2017xas, deSanti:2022jlq}, but also to better recover properties related to the \emph{average} label distribution.

\begin{figure}
    \centering
    \includegraphics[width=\columnwidth]{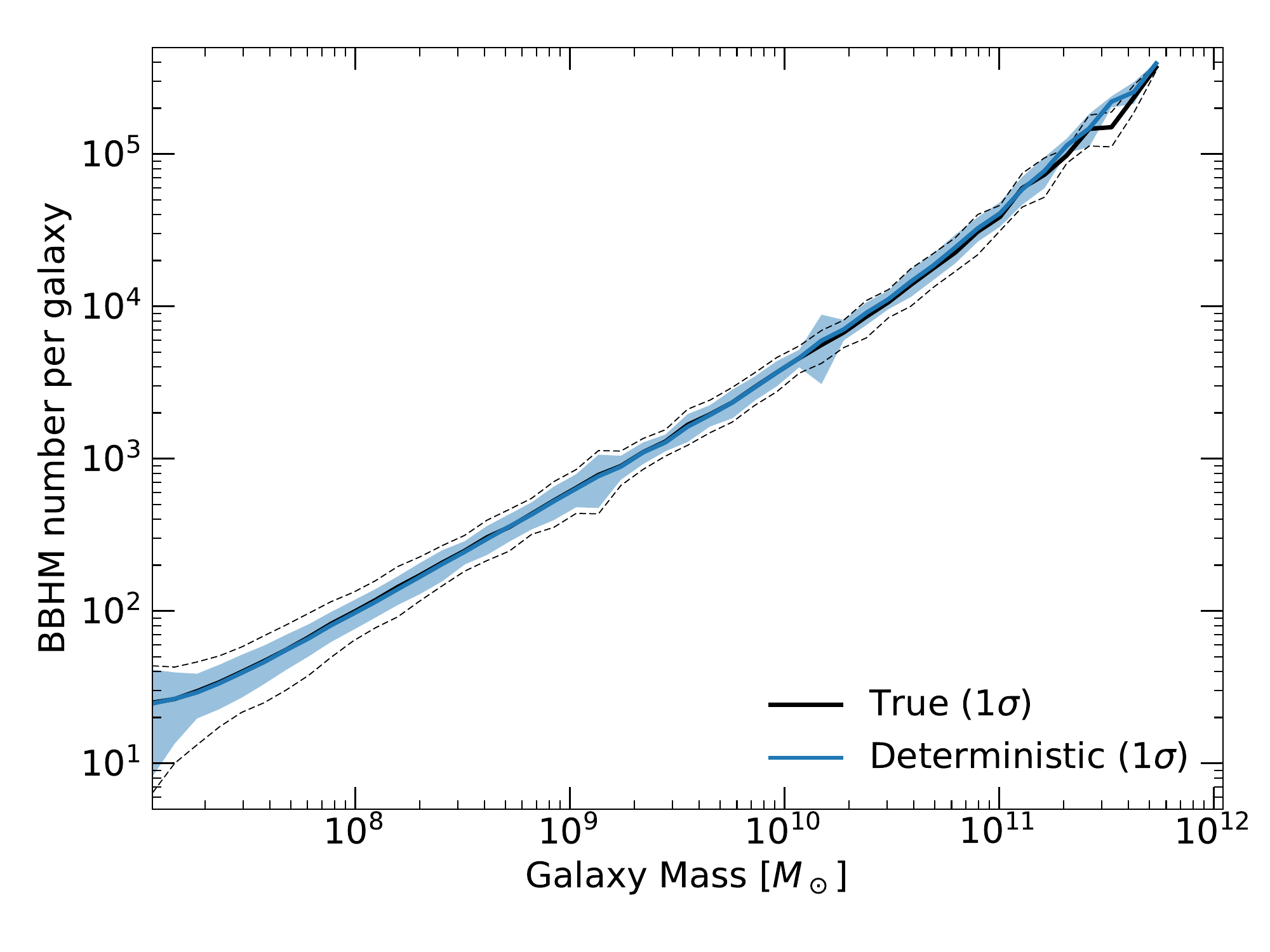}
    \includegraphics[width=\columnwidth]{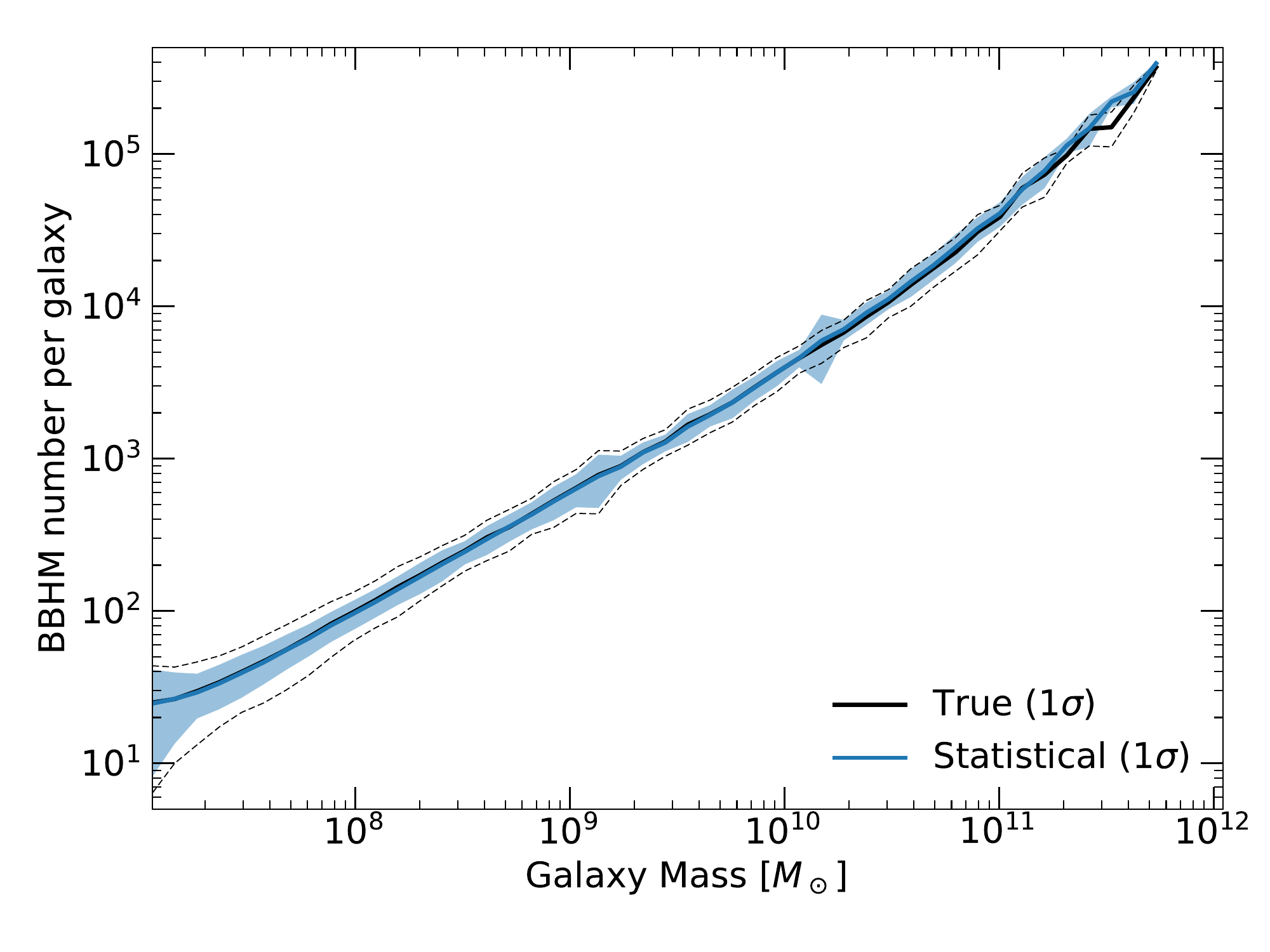}
    \caption{BBHM number distribution as a function of the host galaxy mass. The black line describes the ``true" distribution from the test set, while the blue line describes the NN result. The top panel refers to the deterministic architecture, while the bottom panel relates with the PDF (statistical) one.}
    \label{fig:BHM_vs_galaxy_mass}
\end{figure}

\begin{figure}
    \centering
    \includegraphics[width=\columnwidth]{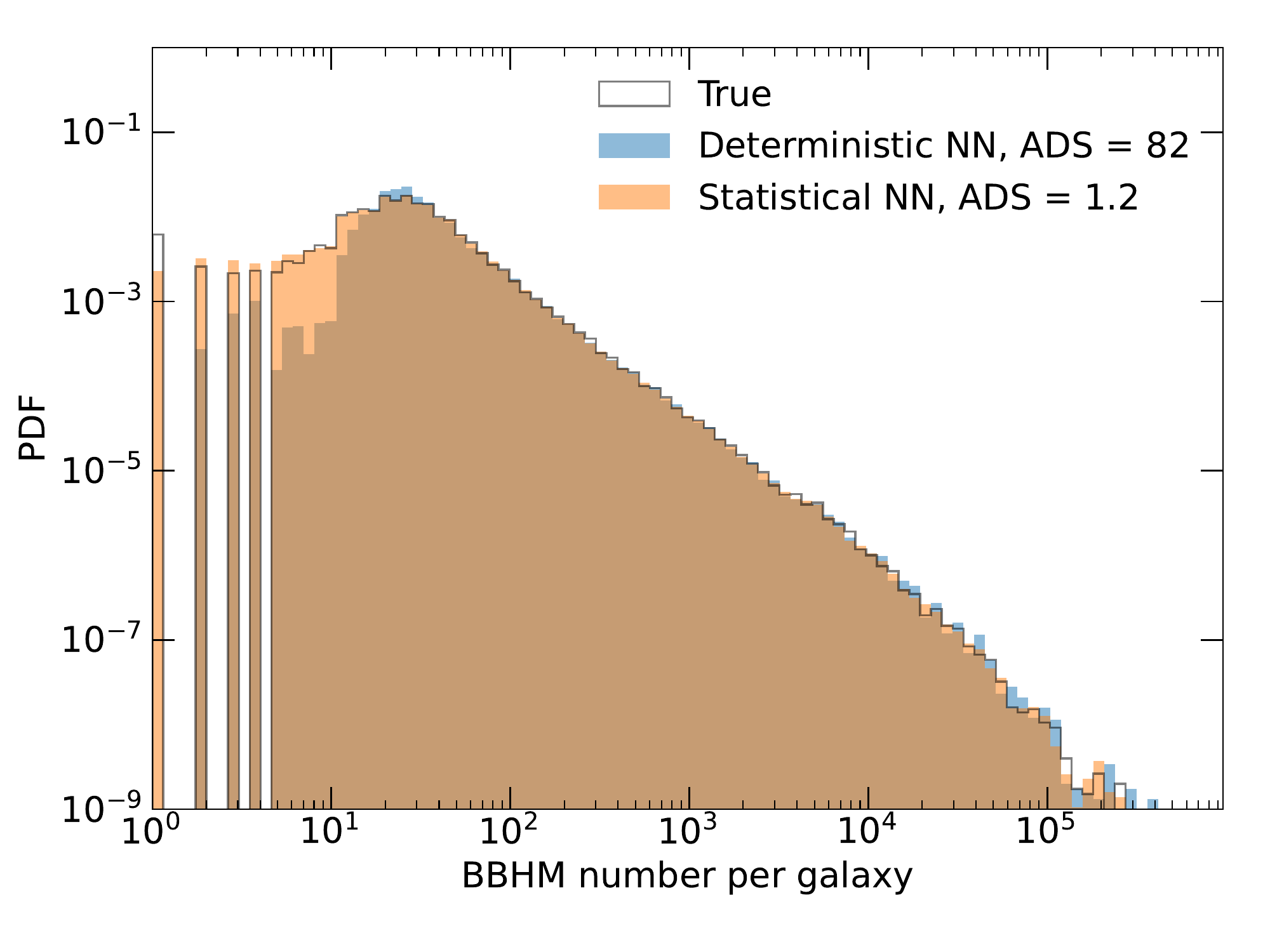}
    \caption{BBHM number PDF per galaxy, marginalized over all the host galaxy and halo properties. The deterministic architecture fails in recovering the ``true" distribution at the low and high number ends of the distribution.}
    \label{fig:BHM_number}
\end{figure}

First, we can test the NN ability to reproduce a BBHM distribution that is statistically equivalent to the test set one, dubbed \emph{true} in the following.
In the two panels of Fig.~\ref{fig:BHM_vs_galaxy_mass} we compare how well the two architectures described in section~\ref{sec:ML} reproduce the true distribution 
as a function of the host galaxy mass.
The host galaxy is singled out among all features because it correlates the most with the target BBHM distribution (see the discussion in the following and Fig.~\ref{fig:AD_feature_importance}).
In each host galaxy mass bin (we use 50 logarithmically spaced bins), the average number of events is reproduced successfully both using the deterministic and the statistical network. However, in most bins, the former generates a distribution which underestimates the true scatter in the bin.
In practice, galaxies that host an extreme (either high or low) number of events are more likely to be populated by the deterministic NN with a number of BBHM closer to the average.
The statistical network, instead, increases the variance of the proposal distribution from which the predictions are drawn when the population of events falls in a broad spectrum. This means that, on a galaxy-by-galaxy basis, the prediction of this NN might differ substantially from the true one, but in such a way that the whole distribution is more closely followed.
The same behaviour appears in the global distribution of BBHM per galaxy shown in Fig.~\ref{fig:BHM_number}.
To provide a quantitative metric of this intuition we use the $k$-sample Anderson Darling test \citep{10.2307/2288805}, 
whose score (ADS) quantifies how close the distributions underlying two samples are to each other. In other words, the ADS is related to the likelihood that the two samples are drawn from the same distribution.
In the following we will always calculate the ADS score of a sample against the true one.
While the true sample and the one predicted from the deterministic network are fixed, the samples generated by the statistical network change depending on the seed.
While the general trends of all the statistical-network-generated samples are consistent between runs, their ADS scores varies.
To produce consistent results, we produce 100 realizations, all based on the same galaxy test set, and we concatenate them to calculate the ADS.
For the deterministic network we find ADS = 80, making the significance of the two sets following the same distribution vanishingly small.
On the other hand, for the statistical network we calculate ADS = 1.2, thus we cannot discern it from the true distribution at <95\% C.L.

\begin{figure}
    \centering
    \includegraphics[width=\columnwidth]{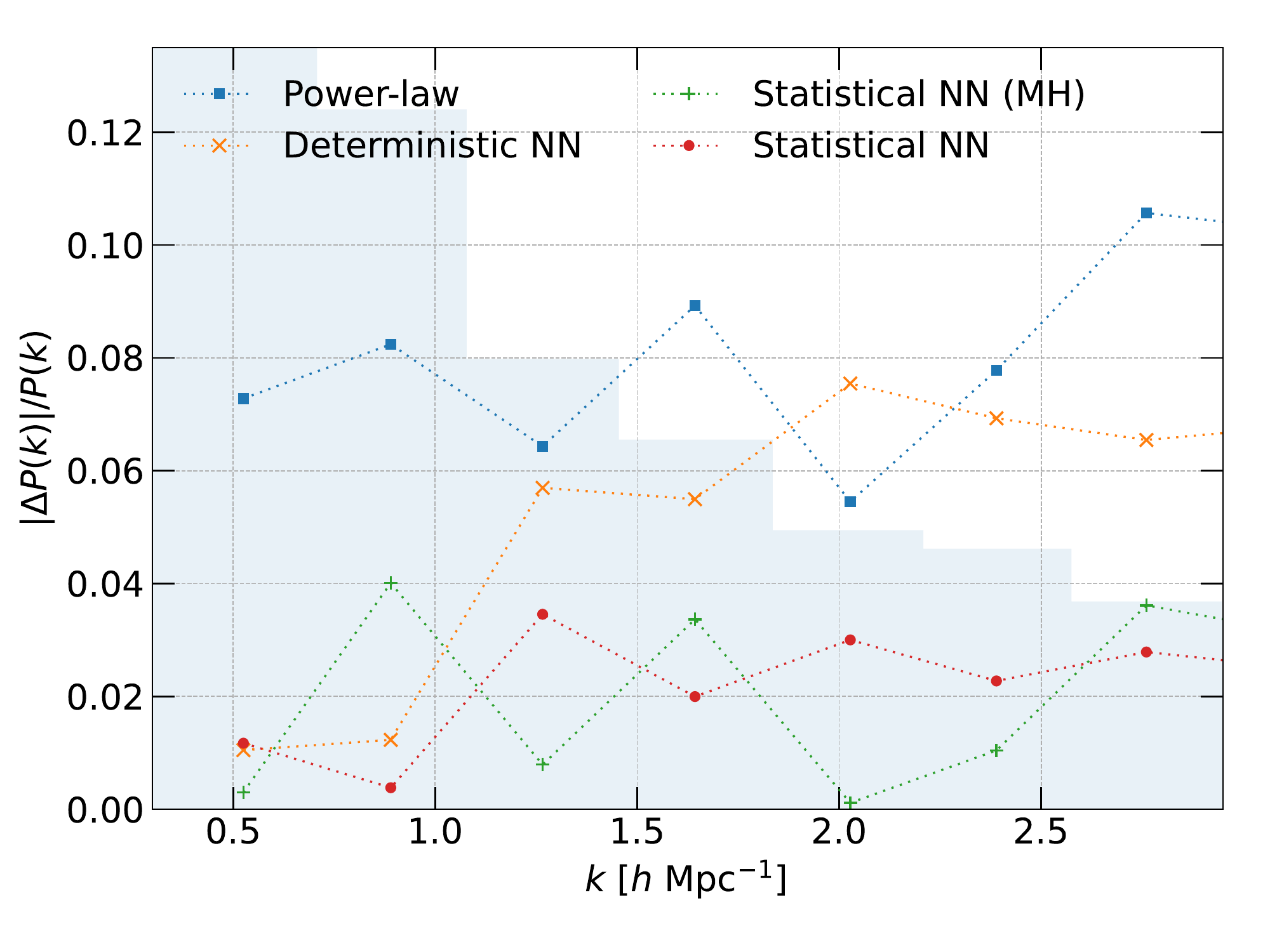}
    \caption{Fractional difference between the N-body power spectrum and different population methods.
    A simple power-law (blue squares) is already capturing the power spectrum at 10\% level.
    Using the statistical NN approach (red dots) improves the emulation compared to methods not describing the spread of the population (orange x's). Basing the training on just the host halo mass (green crosses) does not compromise the results.}
    \label{fig:NN_Pk_differences}
\end{figure}

Second, we are interested in analysing the difference in performance at the power spectrum level. In Fig.~\ref{fig:NN_Pk_differences} we show, as a function of wavemodes, the fractional difference between the merger power spectrum extracted from EAGLE25 (our test sample) and the merger power spectrum of the same simulation painted with the different NN architectures.
We highlight how (due to the smaller box size) the scales represented here only marginally coincide with those used in our bias analysis, and with the scales relevant for Cosmology.
However, in this context the power spectrum ought to be interpreted as a generic signal that we can use to test our painting scheme, without venturing into any claim about what cosmological information could be extracted from it.
While the deterministic NN architecture already achieves a $\leq 5\%$ agreement, which on large scales is negligible with respect to sample variance, 
the statistical NN architecture leads to a factor $\approx 2$ improvement.

Having demonstrated the better performance of the statistical network, we will from now on only use this architecture.

%%%%%%%%%%%%%%%%%%%%%%%%%%%%%%%%%%%%%%%%%%%%%%%%%%%%%
\subsection{Feature importance}
%%%%%%%%%%%%%%%%%%%%%%%%%%%%%%%%%%%%%%%%%%%%%%%%%%%%%
We test the relative importance of features by re-training the NN with different sets of features, comparing each network performance.
Since most of the features are strongly correlated, this is a safer procedure than simply calculating the permutation importance, even if more costly. In the remainder we use the Spearman's rank correlation coefficient $\rho$ as a measure of correlation  \citep[see e.g.][]{daniel1990applied}.
We notice that random-forest-based conditional permutation importance \citep{Strobl_2008} would be a valid alternative, that we will consider in future work, to assess every feature separately.
Since, on the contrary to the the previous section, the loss function employed by the network is now fixed, we can use as our performance metric not only the ADS, but also the likelihood used as loss function itself. While the ADS is sensitive to the global distribution (regardless of, e.g., host galaxy mass, SFR, metallicity, etc.), the likelihood instead is local in feature space, providing complementary information. 

We start by assessing how important the information about the progenitors is. Training the statistical NN on the parameters pertaining to the snapshot where the merger happen, we calculate $\text{ADS}=46$. This is to be expected \citep{Scelfo:2020jyw, Bellomo:2021mer} as what is relevant is the physical state of the host at the time of the binary formation. After the formation, the evolution of the binary, until its eventual merging is only slightly influenced by changes in the surrounding environment.

\begin{figure}
    \centering
    \includegraphics[width=\columnwidth]{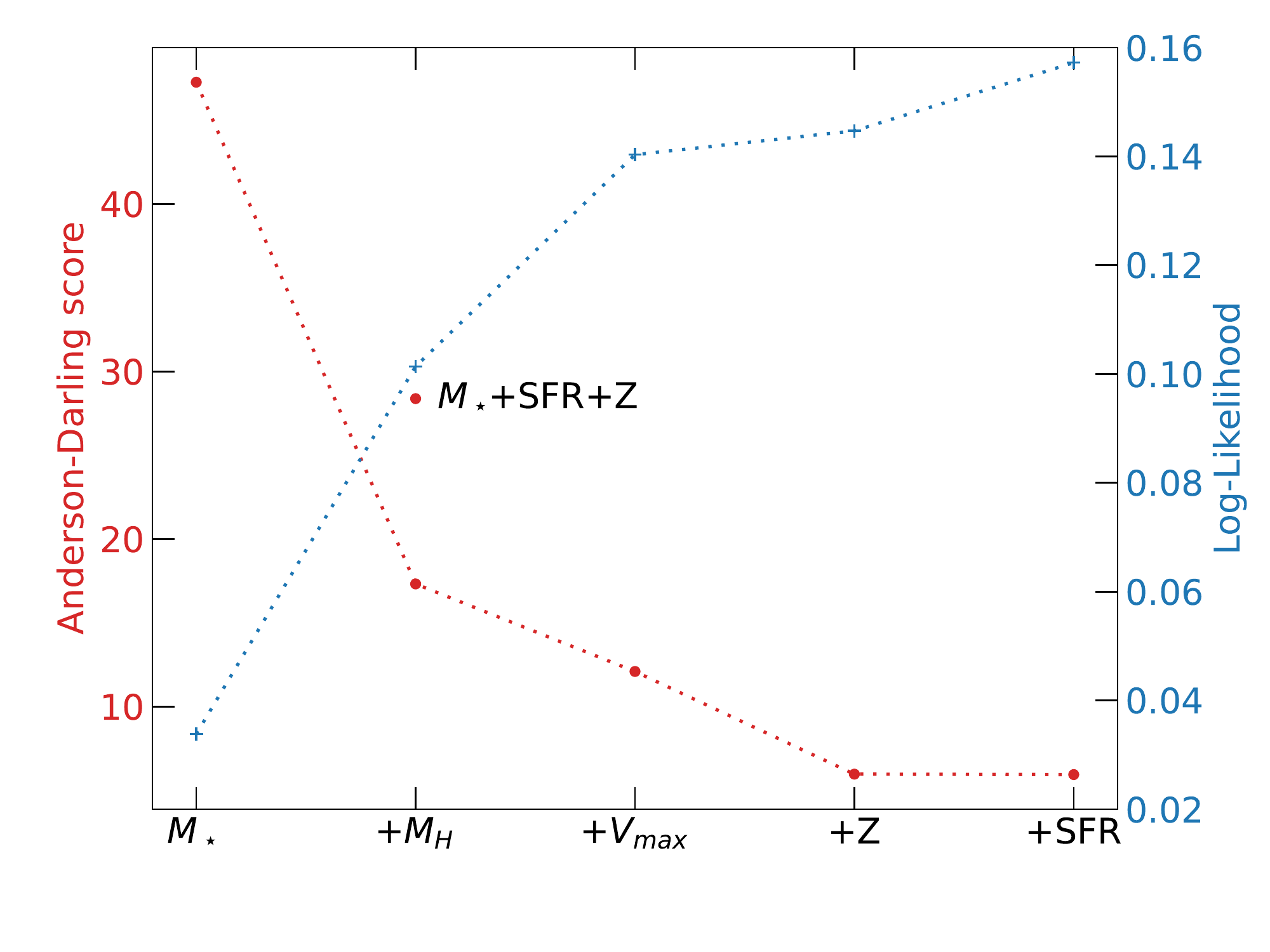}
    \caption{Anderson-Darling score and log-likelihood for different (incremental) sets of features. Each set includes all labels on its left plus the labels indicated on the axis.
    }
    \label{fig:AD_feature_importance}
\end{figure}

Having assessed that information about the host progenitors is crucial, we turn our attention to 5 different (groups of) parameters: $M_\star$, $M_H$, $\psi$, $Z$, $V_\text{max}$.
In each group, say $M_\star$, we include the value of that parameter for the host, and for the progenitors of the host at z=1, 2, 3, say $(M_\star(z=0), M_\star(z=1), M_\star(z=2), M_\star(z=3))$.
Different determinations of the halo mass are not included as $M_H$ strongly correlates with them at each redshift ($\rho \geq 0.95$), acting as a reliable proxy.
The same goes for the parametes related to the kinetic state of the galaxy and $V_\text{max}$ ($|\rho| \geq 0.92$).
An exception is made for $Z$ and $\psi$ which are both included despite their correlation ($\rho \geq 0.84$), because of their physical importance and relevance usually addressed to them in the literature.
We train 5 networks, each of which only receives as input one single group of variables, and select the best according to ADS and likelihood. In all cases investigated here the two metrics produced the same rankings.
The most relevant feature is fixed and 4 networks are trained receiving in input that one and another feature, and the process is repeated.
The progression is shown in Fig.~\ref{fig:AD_feature_importance}.
The host galaxy mass is the most relevant feature to determine the BBHM number distribution (similar to what discussed in A19 and \citealt{2021MNRAS.502.4877S}), together with the mass of its first-progenitors at different snapshots.
We find that the mass of the host halo and $V_\text{max}$, which we remind is a proxy of the kinetic state of the galaxy, are also extremely relevant.
Deeming this worth of further investigation we test if this is driven by $M_H$, $V_\text{max}$ and $M_\star$ acting in concert as a proxy of the relevant information contained in $\psi$ and $Z$.
If this were the case, a network trained with $\psi$, Z, and $M_\star$ would perform better than one trained with only $M_\star$, $M_H$, and $V_\text{max}$.
However, as shown in Fig. \ref{fig:AD_feature_importance}, this does not hold.
Therefore, we confirm that the large-scale environment in which the host galaxy forms---meaning its position in the cosmic web, if it is a central or satellite galaxy, etc.---drives the amount of BBHM. This is akin to how the large-scale environment of an halo influences the number and type of galaxies it hosts \citep{10.1093/mnras/sty2110}.

$Z$ and $\psi$ rank as the least important variables, and the addition of one over the other brings very little additional information. This is not to say that they are physically irrelevant, but rather that since they are extremely correlated to each other, and quite correlated to $M_\star$ they are partially redundant. In fact, this is specifically exploited in e.g. S20 through the use of the AMR.

As for the power spectrum, we find it to be a quite forgiving observable. In fact, even assigning to each galaxy the number of BBHM using the cubic polynomial interpolant (i.e. setting all the fractional residual to 1), the error on the power spectrum is $\lesssim 10\%$, while training the statistical network just with the host galaxy mass as input, the deviation is $\lesssim 4\%$ (Fig. \ref{fig:NN_Pk_differences}). Training the network on the full set of parameters does not lead to an improvement.
Our analysis highlights a difference in the relative importance of the additional parameters to fully characterize the power spectrum and BBHM distribution. 
The existence of a relation that adequately links the sole mass of the halo to the BBHM power spectrum, while incredibly useful to populate relatively cheap N-body simulations, is not helpful in understanding the underlying physics. If one's goal is the latter, and they wish to do so by building an HOD, the use of the BBHM distribution cannot be bypassed. In this sense, our analysis informs how bias models shall be built.

%%%%%%%%%%%%%%%%%%%%%%%%%%%%%%%%%%%%%%%%%%%%%%%%%%%%%
\subsection{Apply the NN to paint BBHM}
%%%%%%%%%%%%%%%%%%%%%%%%%%%%%%%%%%%%%%%%%%%%%%%%%%%%%
The machinery demonstrated in the previous sections can in principle be readily used to paint hydrodynamic simulations. The resulting BBHM catalogues are statistically equivalent to those obtained with the procedure described in section~\ref{sec:sims}, but can be produced in a few tens of seconds using a single GPU.
However, a few caveats are due.
By virtue of a separation of scale argument, we can expect that the formation of BH binaries and their subsequent evolution should mostly be influenced by their galactic environment and not from the cosmological background. Thus, changes in the cosmological parameters should have negligible effect on the BBHM content of any given galaxy, so we can expect that the NN can be safely generalized to other simulations, even when they assume (at least slightly) different cosmologies. However, this hypothesis ought to be verified prior to any application.
Moreover, it is known \citep[e.g.][]{Villaescusa-Navarro:2020rxg} how different hydrodynamical models produce qualitatively different simulations, and as such the network might fail to populate simulations which are not created with the same algorithm, as seen in e.g. \citet{Villaescusa-Navarro:2021pkb}.

%%%%%%%%%%%%%%%%%%%%%%%%%%%%%%%%%%%%%%%%%%%%%%%%%%%%%
%%%%%%%%%%%%%%%%%%%%%%%%%%%%%%%%%%%%%%%%%%%%%%%%%%%%%
\section{Conclusions}\label{sec:conclusion}
%%%%%%%%%%%%%%%%%%%%%%%%%%%%%%%%%%%%%%%%%%%%%%%%%%%%%
%%%%%%%%%%%%%%%%%%%%%%%%%%%%%%%%%%%%%%%%%%%%%%%%%%%%%

With the fourth LIGO--Virgo--KAGRA observational campaign approaching \citep{2020LRR....23....3A} and 3rd generation detectors planned for the next decade \citep{2021arXiv211106990K,2023arXiv230315923B}, it is becoming more and more important to build accurate models to understand the properties of GW sources. These include their redshift distribution and clustering properties, which, in the case of astrophysical binary mergers, ultimately depend on the way their progenitors form according to the properties of their host halo and galaxy.

In this work, we focused on the study of the clustering of astrophysical binary black hole mergers (BBHMs). We relied on cosmological, hydrodynamical simulations~\citep{artale_2019,artale2020} to estimate the cosmological linear bias of such events at different redshifts. On one side, we developed a fully-numerical estimator (publicly available at: \url{https://github.com/MatPeron/powerbias}) which can be applied to mock or real GW catalogues. Because of the small scales accessible by the size of the simulation adopted, we needed to account for tree-level and 1-loop contributions to the power spectrum and to marginalized over nuisance, higher order bias parameters. Our algorithm was tested by comparing the theoretical DM power spectrum and halo bias with their estimates on larger N-body simulations, in which linear scales can be accessed and leading order truncation can be performed. On the other side, we implemented a semi-analytical HOD-based approach~\citep{Libanore:2020fim,Libanore:2021jqv} to estimate the BBHM bias as a function of the properties of their host galaxies. The method was calibrated on the same simulations as the fully-numerical estimator and the two approaches show internal consistency.
We also compared our results with an independent semi-analytical model~\citep{Scelfo:2020jyw} based on merger rate-weighting of the host galaxy bias.
Our finding is that the different approaches are consistent, provided that the underlying assumptions on the astrophysical modelling and the properties of the host halos and galaxies are in agreement. While this seems to be obvious, the important message is that, if we want to use the GW bias as a tool to perform precision cosmological studies, we first of all need to understand how to properly account for uncertainties in the formation channels and host properties.
As a first step in this direction, we implemented a neural network aimed at studying the relative relevance of the host parameters with respect to the BBHM distribution and bias. While for the former it is important to account for the host galaxy mass, as well as the host halo mass and the galaxy properties at the time of the binary formation, the latter (which in the end depends on how the power spectrum behaves) seems to suppress the relevance of parameters other than the host galaxy or host halo mass.
Our neural network can also be applied to paint BBHM on N-body or cosmological, hydrodynamical simulations quickly and with a low computational cost. To improve its reliability, the neural network should be tested on simulations accounting for different cosmologies and astrophysical models, which are not available yet in the context of GW sources.

To conclude, with our work we compared the different GW bias prescriptions available on the market and we provided bias estimates based on the state-of-the-art situation in BBHM simulations. Our results in Tab.~\ref{tab:MCMC_b1} and the fit in Eq.~\eqref{eq:fit} can be used as priors to perform cosmological studies. Improving these findings will be more and more important as the number of GW observations will increase. Once observed GW catalogues will allow to perform statistical studies up to cosmological distances, the clustering will be the perfect tool to perform a wide variety of tests, from the progenitor formation channels (here we only dealt with astrophysical binary black holes, while primordial binary black holes, if they exist, should have a completely different bias~\citep{raccanelli_2016,Scelfo:2020jyw,Libanoreinprep}), to modified gravity models~\citep{Scelfo:2022lsx,Balaudo:2022znx} or redshift-space distortion effects~\citep{Mukherjee:2018ebj,Mukherjee:2020hyn}. To achieve this goal, the next challenges will be to understand how to access larger scales with the mock catalogues and how to include selection and observational effects, so to realize mock data-sets as close as possible to the real ones. Once these will be available, it will be possible to updated the tools we realized in order to make them applicable to real estimates.

%-------------------------------------------------------------------
\section*{Data Availability} 
The self-consistent algorithm developed to estimate the (more than tree-level) power spectrum and analyse the clustering properties of {\sc eagle}+\mobse{} data-set is publicly available at \url{https://github.com/MatPeron/powerbias}.
The trained neural network models are available upon reasonable request. \mobse{} is an open source code available at \url{https://gitlab.com/micmap/mobse_open}.
%-------------------------------------------------------------------

%%%%%%%%%%%%%%%%%%%%%%%%%%%%%%%%%%%%%%%%%%%%%%%%%%%%%
%%%%%%%%%%%%%%%%%%%%%%%%%%%%%%%%%%%%%%%%%%%%%%%%%%%%%
\section*{Acknowledgements}
We acknowledge M.~Mapelli support in realizing the mock data-set and interpreting the results. 
We thank D.~Karagiannis, A.~Ricciardone and E.~D.~Kovetz for useful discussions. 
We acknowledge support by the project "Combining Cosmic Microwave Background and Large Scale Structure data: an Integrated Approach for Addressing Fundamental Questions in Cosmology", funded by the MIUR Progetti di Ricerca di Rilevante Interesse Nazionale (PRIN) Bando 2017 - grant 2017YJYZAH. AR acknowledges support from PRIN-MIUR~2020 METE, under contract no. 2020KB33TP. SL~acknowledges support by the Fondazione Ing.\ Aldo Gini and the Azrieli Foundation.
The computational results presented here have been achieved (in part) using the LEO HPC infrastructure of the University of Innsbruck.
%%%%%%%%%%%%%%%%%%%%%%%%%%%%%%%%%%%%%%%%%%%%%%%%%%%%%
%%%%%%%%%%%%%%%%%%%%%%%%%%%%%%%%%%%%%%%%%%%%%%%%%%%%%



\bibliographystyle{mnras}
\bibliography{biblio} % if your bibtex file is called example.bib

%%%%%%%%%%%%%%%%%%%%%%%%%%%%%%%%%%%%%%%%%%%%%%%%%%

%%%%%%%%%%%%%%%%% APPENDICES %%%%%%%%%%%%%%%%%%%%%

\appendix

%%%%%%%%%%%%%%%%%%%%%%%%%%%%%%%%%%%%%%%%%%%%%%%%%%%%%
%%%%%%%%%%%%%%%%%%%%%%%%%%%%%%%%%%%%%%%%%%%%%%%%%%%%%
\section{1-loop Bias Operators}\label{sec:bias_ops}

In Eqs.~\eqref{eq:auto_exp}, \eqref{eq:cross_exp} we define the auto- and cross-power spectrum expansions adopted in the {\sc EFTofLSS} estimator. The tree-level and 1-loop contributions that enter such equations are defined as in~\citet{Desjacques:2016bnm} and presented here for ease of reference:

\begin{align*}
    F_{\delta^2}(k)&\equiv0\,, \\
    I_{\delta^2}(k)&\equiv2\int_{\boldsymbol{q}}F_2(\boldsymbol{k}-\boldsymbol{q},\,\boldsymbol{k})P_L(k)P_L(|\boldsymbol{k}-\boldsymbol{q}|)\,, \\
    F_{K^2}(k)&\equiv4P_L(k)\int_{\boldsymbol{q}}\sigma^2_{q,\,k-q}F_2(\boldsymbol{k},\,-\boldsymbol{q})P_L(q)\,, \\
    I_{K^2}(k)&\equiv2\int_{\boldsymbol{q}}\left(\sigma^2_{q,\,k-q}+\frac{2}{3}\right)F_2(\boldsymbol{k}-\boldsymbol{q},\,\boldsymbol{k})P_L(k)P_L(|\boldsymbol{k}-\boldsymbol{q}|)\,, \\
    F_{td}(k)&\equiv\frac{8}{5}P_L(k)\int_{\boldsymbol{q}}\sigma^2_{q,\,k-q}F_2(\boldsymbol{k},-\boldsymbol{q})P_L(q)\,, \\
    I_{td}(k)&\equiv0\,,\\
    I_{\delta^2\delta^2}(k)&\equiv2\int_{\boldsymbol{q}}P_L(q)P_L(|\boldsymbol{k}-\boldsymbol{q}|)\,, \\
    I_{\delta^2K^2}(k)&\equiv2\int_{\boldsymbol{q}}\left(\sigma^2_{q,\,k-q}+\frac{2}{3}\right)P_L(k)P_L(|\boldsymbol{k}-\boldsymbol{q}|)\,, \\
    I_{K^2K^2}(k)&\equiv2\int_{\boldsymbol{q}}\left(\sigma^2_{q,\,k-q}+\frac{2}{3}\right)^2P_L(k)P_L(|\boldsymbol{k}-\boldsymbol{q}|)\,,
\end{align*}
where $b_{\delta^2}\equiv b_{2,t}/2$, $\sigma^2_{q,\,k}=(\boldsymbol{q}\cdot\boldsymbol{k}/qk)^2-1$.

%%%%%%%%%%%%%%%%%%%%%%%%%%%%%%%%%%%%%%%%%%%%%%%%%%%%%
%%%%%%%%%%%%%%%%%%%%%%%%%%%%%%%%%%%%%%%%%%%%%%%%%%%%%
%%%%%%%%%%%%%%%%%%%%%%%%%%%%%%%%%%%%%%%%%%%%%%%%%%%%%
%%%%%%%%%%%%%%%%%%%%%%%%%%%%%%%%%%%%%%%%%%%%%%%%%%%%%

\section{Power spectrum estimator}
\label{sec:estimators}

The snapshots and catalogues we used in our analysis contain 
$N_p$ point-like particles (DM or tracer) distributed inside a cubic box of size $L$. Each particle is localized within the box by $\boldsymbol{r} = (x,\,y,\,z)$, where the origin $(0,\,0,\,0)$ is at one of the box vertices. We define the density as $\rho(\boldsymbol{r})\equiv m\sum_{i=1}^{N_p}\delta_D(\boldsymbol{r}-\boldsymbol{r}_i)$ and the density contrast as \begin{equation}\label{eq:appendix_delta}
\delta(\boldsymbol{r})  \equiv\frac{\rho(\boldsymbol{r})}{\bar{\rho}}-1  =\frac{\sum_{i=1}^{N_p}\delta_D(\boldsymbol{r}-\boldsymbol{r}_i)}{\bar{n}}-1,
\end{equation}
where $\bar{\rho}=m\bar{n}=mN_p/L^3$ is the spatially averaged density within the box. On a theoretical level, the power spectrum is then defined by taking the Fourier transform $\delta(\boldsymbol{k})$, computing its square modulus $|\delta(\boldsymbol{k})|^2=\delta(\boldsymbol{k})\delta^*(\boldsymbol{k})$, and averaging the resulting modes inside $k$ bins.
This procedure assumes that the fair sample hypothesis holds.

Since we are dealing with discrete data-sets, 
the density field is unevenly sampled in space. Therefore, before applying the fast Fourier transform (FFT) %techniques 
we need to smear $\delta(\boldsymbol{r})$ onto a regular grid within the box by convolving it with a proper filter~\citep{tegmark_1998}. In this way, we interpolate $\tilde{\delta}(\boldsymbol{r})$; to recover $\delta(\boldsymbol{k})$, one then has to deconvolve the filter, which in Fourier space is simply done by dividing the Fourier transform of the filter to $\tilde{\delta}(\boldsymbol{k})$.

With this basic recipe in mind, we can discuss in more detail the steps we take in order to arrive to an estimator $\tilde{P}(k)$. 
\begin{enumerate}
    \item The filter function $W(\boldsymbol{r})$ we adopt is the standard Cloud-In-Cell (CIC) weighting scheme \citep[see equation A.2 of][]{CHANIOTIS2004253}, which smears a single particle across a cube of eight neighbouring cells. 
    Chosen %a filter function $W(\boldsymbol{r})$ and 
    a number of grid cells per side $N_g$, we use it to compute the weights $W(\boldsymbol{r}_g)$ for each particle at each grid's node position $\boldsymbol{r}_g$. 
    
    The weights are then summed across the entire catalogue to estimate the number density field $\tilde{n}(\boldsymbol{r}_g)=\tilde{\rho}(\boldsymbol{r}_g)/m$.
    \item $\tilde{\delta}(\boldsymbol{r}_g)$ is found according to equation \ref{eq:appendix_delta}.
    \item $\tilde{\delta}(\boldsymbol{k})$ is given by performing the FFT of $\tilde{\delta}(\boldsymbol{r}_g)$.
    \item The Fourier transform of the CIC filter function $W(\boldsymbol{r})$ is computed analytically and divided to $\tilde{\delta}(\boldsymbol{k})$ in order to recover the actual Fourier transform of the field, $\delta(\boldsymbol{k})$.
    \item The proper normalizations given by the FFT and the discretization of the Dirac deltas are applied to $\delta(\boldsymbol{k})$.
    \item We compute $\tilde{P}(k)$ by averaging $|\delta(\boldsymbol{k})|^2$ inside bins in Fourier space. Bins are defined inside the interval $[k_{\rm f},\,k_{\rm Nyq})$ and are $k_{\rm f}$-wide, where $k_{\rm f}=2\pi/L$ is the fundamental mode, while $k_{\rm Nyq}=\pi N_g/L$ is the Nyquist mode, i.e.,~the maximum mode supported by the grid. 
    \item We compute error bars for the estimator by taking the square root of the Gaussian variance, $\sqrt{\mathrm{Var}(\tilde{P}(k))|_G}=\sqrt{2/N_k}\tilde{P}(k)$, where $N_k$ is the number of independent Fourier modes within the bin \citep{Scoccimarro:1999kp}. We neglect non-Gaussian contributions related with the four-point function of $\delta(\boldsymbol{k})$ (namely, the trispectrum) since they are only significant at even smaller scales than the ones we consider~\citep{Mohammed:2016sre}.
\end{enumerate}

Since we are computing $\tilde{P}(k)$ from a discrete collection of particles, shot-noise arises from their auto-correlations. We can assume that the particles are Poisson sampled from the underlying density field; under this assumption, the shot noise contribution is equal to $\bar{n}^{-1}$ \citep[see appendix A of][]{Feldman:1993ky}. In our analysis, we subtract it before making any measurement of the bias parameters.

Another consideration to make is that by choosing to smear the particles onto a regular grid, we are effectively shifting all the power associated to the modes $k>k_{\rm Nyq}$ inside the range of modes supported by the grid. This effect is known as aliasing and it causes a spurious increase in power at modes  $k\gtrsim k_{\rm Nyq}/2$ \citep{colombi_2009}. Aliasing can be alleviated \citep{sefusatti_2016} or outright removed under certain assumptions on the shape of $P(k)$ \citep{jing_2005}. We implemented the interlacing approach from \citet{sefusatti_2016} to deal with this effect (see Fig.~\ref{fig:quijoteDM_vs_CAMBnonlinear}).
It is important to note that interlacing, while useful if seeking accurate estimation from large scale simulations, is not necessary for small-side simulations such as \textsc{eagle}. In fact, the range of modes where the perturbative prescription of Eq.~\eqref{eq:bias_exp} is valid is very close to the fundamental mode, where aliasing has no effect. For this reason, we only use interlacing to analyse~\textsc{Quijote} data in the validation test performed in appendix~\ref{sec:validation}.

The procedure we depicted up to now is optimal to estimate the power spectrum on ideal, mock data-sets; with the prospect of applying the same methodology to real data, however, selection effects should be taken into account.
For this reason, in our algorithm we also included the Feldman-Kaiser-Peacock (FKP) estimator \citep{Feldman:1993ky}, in which the data-set is compared with a synthetic random catalogue, in order to optimize the variance. In our case, this catalogue is generated by uniform sampling a set of particles inside the same box in which the simulation has been run.
The implementation follows the same general scheme presented above, with some extra-steps related with the synthetic catalogue and the optimal weighting that characterizes the FKP estimator. In particular, step (i) in our algorithm is applied twice, once on the mock catalogue and another on the synthetic, random catalogue. Then, before step (ii), we define the optimal weights $w(\boldsymbol{r})=\left[\bar{n}(\boldsymbol{r})P_0(k)+1\right]^{-1}$ from a initial guess of the power spectrum $P_0(k)=\mathrm{const}$. These are used to compute the weighted number density field $F(\boldsymbol{r})$:
\begin{equation}
    F(\boldsymbol{r})=\frac{w(\boldsymbol{r})\left(n(\boldsymbol{r})-\alpha n_s(\boldsymbol{r})\right)}{\sqrt{\int \mathrm{d}^3r\,w(\boldsymbol{r})^2\bar{n}(\boldsymbol{r})^2}},
\end{equation}
where the $s$ subscript refers to %a quantity computed from 
the synthetic random catalogue, while $\alpha\lesssim0.1$ is the ratio between the density of the data catalogue and the synthetic, random one. 
The algorithm then proceeds %In the following steps of the power spectrum estimation, 
using $F(\boldsymbol{r})$ in place of $\delta(\boldsymbol{r})$.
The presence of $P_0(\boldsymbol{k})$ in $w(\boldsymbol{r})$ implies that the weights can be optimized iteratively by using the resulting $\tilde{P}_\mathrm{FKP}(k)$. Therefore, by starting from $P_0(k)=\mathrm{const}$, we get a first estimate $\tilde{P}_\mathrm{FKP}(k)$ which is used to improve the estimate of the optimal weights  bin-by-bin; for each $k$, we then compute a new set of weights $w(\boldsymbol{r})$ using $P_0(k)=\tilde{P}_\mathrm{FKP}(k)$, and obtain a new estimate $\tilde{P}_\mathrm{FKP}(k)$. This is iterated up to some convergence criterion.

\section{Validation tests}\label{sec:validation}

As we anticipated in section~\ref{sec:sims}, the small volume of the {\sc eagle} simulation~\citep{schaye2015} forces us to estimate the power spectrum and bias on small, non-linear scales. Theoretical models in this regime are complicated and depend on large sets of parameters. 
Before using our algorithm on these scales, therefore, we tested it in a ``simpler'' configuration, namely on the large, linear scales provided by the {\sc Quijote} simulation~\citep{villaescusa_2020}.

We now describe the validation tests we performed both on the power spectrum and bias estimators.

\subsection{Power spectrum estimator}

We validate our power spectrum estimator on the dark matter field from the \textsc{Quijote} simulations \citep{villaescusa_2020}. We consider five different realizations with the same cosmological parameters (the fiducial ones). For each, we compute the DM power spectrum for every available redshift snapshot, namely $z\in\{0,\,0.5,\,1,\,2,\,3\}$, interpolating its density contrast field onto a grid of $N_g=512$ cells per side. The resulting 25 power spectra are then averaged across the five realization, obtaining five estimated power spectra, one for each redshift. We then use \texttt{CAMB}\footnote{\url{https://github.com/cmbant/CAMB}.} \citep{Lewis:1999bs,Howlett:2012mh} to compute the theoretical non-linear matter power spectrum given by the halofit model \citep{Mead:2020vgs} for the given cosmology at each redshift. 

\begin{figure}
    \centering
    \includegraphics[width=\columnwidth]{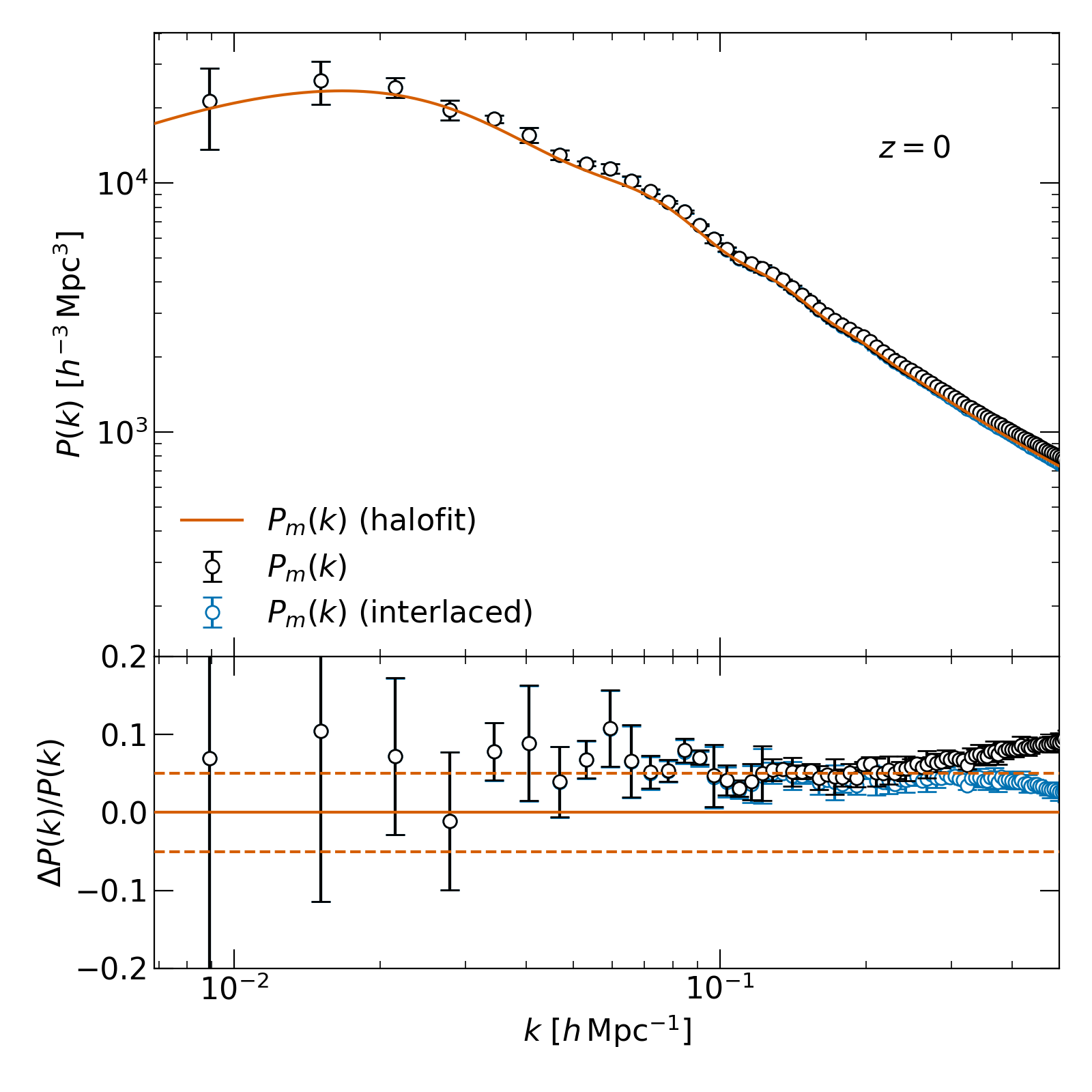}
    \caption{{\sc Quijote} DM power spectrum estimated at $z=0$ with (blue) and without (black) interlacing, compared with $P(k)$ from \texttt{CAMB} (orange line).}
    \label{fig:quijoteDM_vs_CAMBnonlinear}
\end{figure}

Figure \ref{fig:quijoteDM_vs_CAMBnonlinear} compares the estimated and theoretical power spectra at $z=0$.
Up to $k\sim0.2$ $h\,$Mpc$^{-1}$, the power spectrum measured from the simulations agrees within 5\% with the theoretical one. At smaller scales a systematic shift in power is observed, which can be explained by aliasing \citep{sefusatti_2016,jing_2005}. This behavior is observed at all redshifts. Using the interlacing algorithm introduced in~\citet{sefusatti_2016} improves the estimate considerably, pushing the agreement up to $k\sim0.4-0.5$ $h\,$Mpc$^{-1}$.

\subsection{Bias estimator}

We test the MCMC estimator described in section \ref{sec:bias_mock} on DM halos from \textsc{Quijote} simulations \citep{villaescusa_2020}. 

We consider the same five realizations used in the previous section, sorting the halos into six mass bins uniform in log-space. For each mass bin and redshift, we compute the interlaced power spectrum using a grid of $N_g=512$ cells per side to interpolate the density field. We obtain a total of 118 power spectra (we are unable to compute 32 of them due to the absence of halos at high redshifts and masses). Again, we average the different realizations, obtaining 24 estimated power spectra, one for each redshift and mass bin.

Finally, we obtain estimates for the linear bias parameter $b_{1,h}$ by fitting the models of Eqs.~\eqref{eq:auto_exp}-\eqref{eq:cross_exp_limd} to the estimated power spectra.
As in section \ref{sec:bias_mock}, we use a Gaussian likelihood and uniform priors, taking care in setting $b_{1,h}\in[0,\,50]$ to provide enough parameter space for the chains to explore the posterior. We combine auto and cross-power spectra corrected for the shot noise and truncated at a scale $k_\mathrm{NL}$, which we set at 90\% the non-linear threshold scale for matter described in \citet{Smith:2002dz,Fonseca:2019qek}, that is $k_\mathrm{NL}=0.9\cdot0.2(z+1)^{2/(2+n_s)}\,h\,$Mpc$^{-1}$ with $n_s=0.9624$.

\begin{figure}
    \centering
    \includegraphics[width=0.4\textwidth]{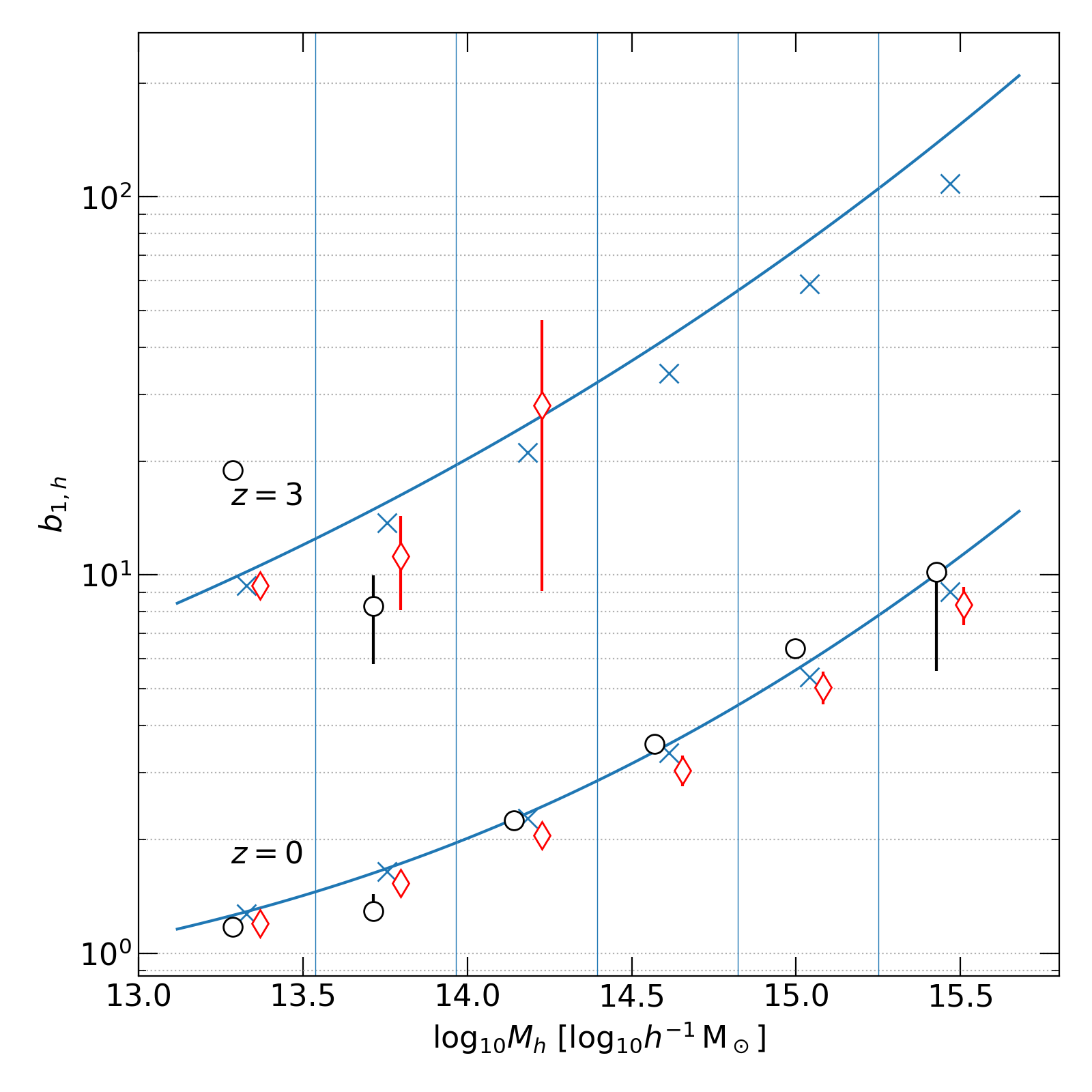}
    \caption{Halo bias at $z = 0$ and $z = 3$ estimated with MCMC based on {\sc EFTofLSS} (black circles), compared with the PBS expectation value weighted by the halo mass function (blue crosses) and the empirical linear bias estimator (red diamonds).}
    \label{fig:quijote_halo_bias_EFTofLSS}
\end{figure}

\begin{figure}
    \centering
    \includegraphics[width=0.4\textwidth]{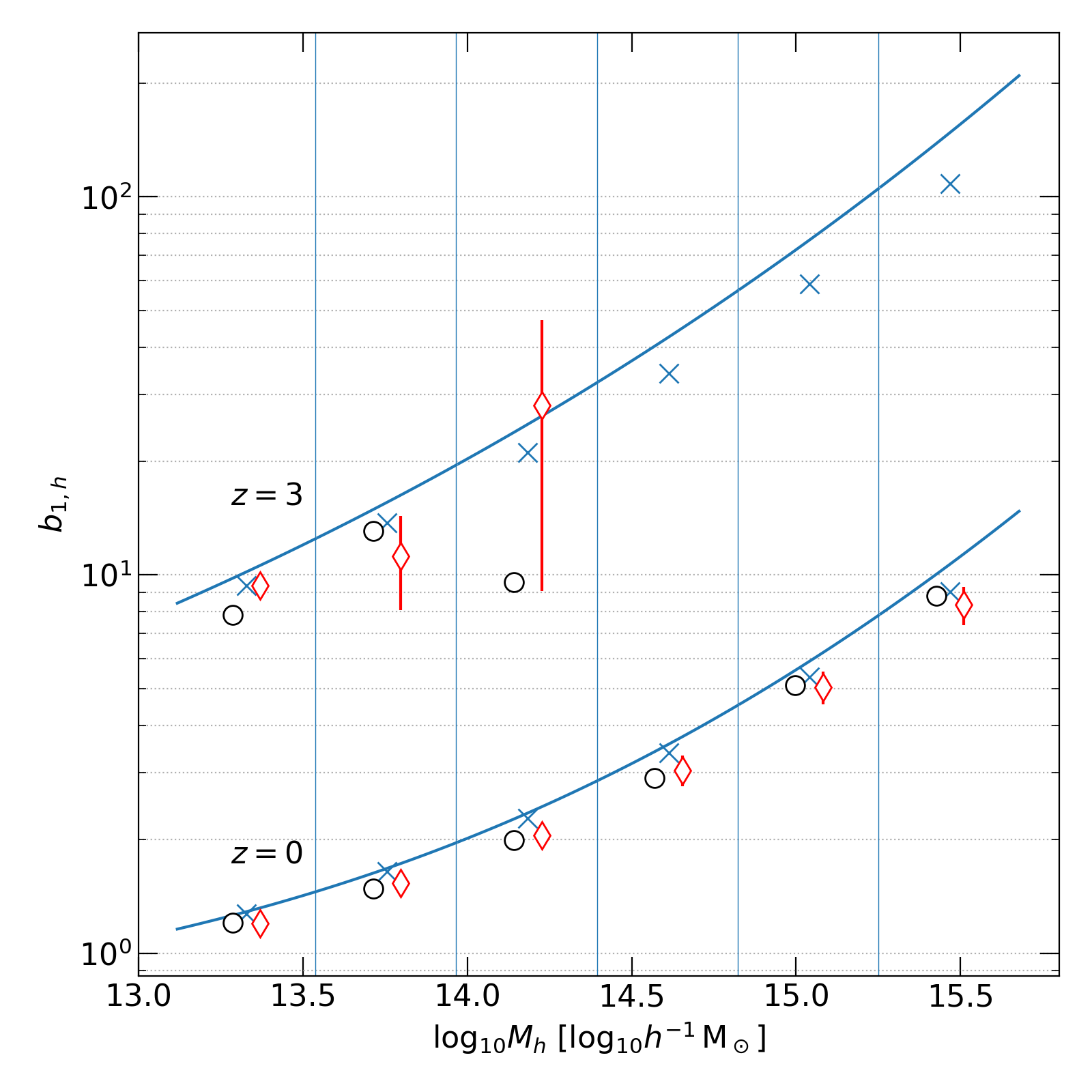}
    \caption{Halo bias at $z = 0$ and $z = 3$ estimated with MCMC based on {\sc LIMD} (black circles), compared with the PBS expectation value weighted by the halo mass function (blue crosses) and the empirical linear bias estimator (red diamonds).}
    \label{fig:quijote_halo_bias_LIMD}
\end{figure}

We compare these estimates with the theoretical expectations of $b_{1,h}$ from the Peak-Background Split (PBS) method, assuming the theoretical halo mass function distribution described by \citet{Tinker:2010my} and computed through the \texttt{hmf} python package. The PBS gives us $b_{1,h}(M_h)$ as a function of the halo mass $M_h$; we use the halo mass function to weigh it inside each mass bin, so to find a single value per bin. Figs.~\ref{fig:quijote_halo_bias_EFTofLSS} and~\ref{fig:quijote_halo_bias_LIMD} compare the {\sc EFTofLSS} and {\sc LIMD} halo bias estimates with the PBS expectation values weighted in each mass bin. Using the Tinker halo mass function to weigh the PBS bias gives results that are in overall agreement with the empirical estimations at linear scales, with some small deviations due to differences between the actual halo mass function of the simulation compared to the Tinker prescription.

As a further test, we estimate the linear bias truncating the %by using only the estimated 
power spectra %truncated 
at $k\sim0.08\,h\,$Mpc$^{-1}$. This leaves us only with the linear scales, where the bias expansions %in section \ref{sec:bias} 
can be truncated at leading order:
\begin{equation}\label{eq:auto_exp_linear}
    P_{tt}(k)=b_{1,t}^2P_L(k)+P_\epsilon,
\end{equation}
\begin{equation}\label{eq:cross_exp_linear}
    P_{tDM}(k)=b_{1,t}P_L(k).
\end{equation}
$b_{1,h}$ can easily be obtained by reversing these equations. These estimates are then averaged together across the various $k$-modes to obtain a final estimate of $b_{1,h}$, with $1\sigma$ errorbars given by their standard deviation. This empirical bias estimate is compared with the MCMC \textsc{EFTofLSS} and \textsc{LIMD} results in Figs. ~\ref{fig:quijote_halo_bias_EFTofLSS} and \ref{fig:quijote_halo_bias_LIMD}.

MCMC estimates obtained with the \textsc{LIMD} model agree really well with the empirical ones, with some noise at high redshifts explainable by the small amount of halos in this range. On the other hand, \textsc{EFTofLSS} results while following the same trend as the other estimator, are noisier, which may be explained by the higher dimensionality of the model itself. It is important to note that, while these results favor the \textsc{LIMD} model over the \textsc{EFTofLSS} one, the binary black hole merger catalogues on which we recover the final result of this paper contain far more objects than the dark matter halos used here, meaning that the measured power spectra will be less noisy. Indeed, section \ref{sec:bias_mock} shows that \textsc{LIMD} and \textsc{EFTofLSS} are in perfect agreement when it comes to binary black hole mergers.

\begin{figure}
    \centering
    \includegraphics[width=0.4\textwidth]{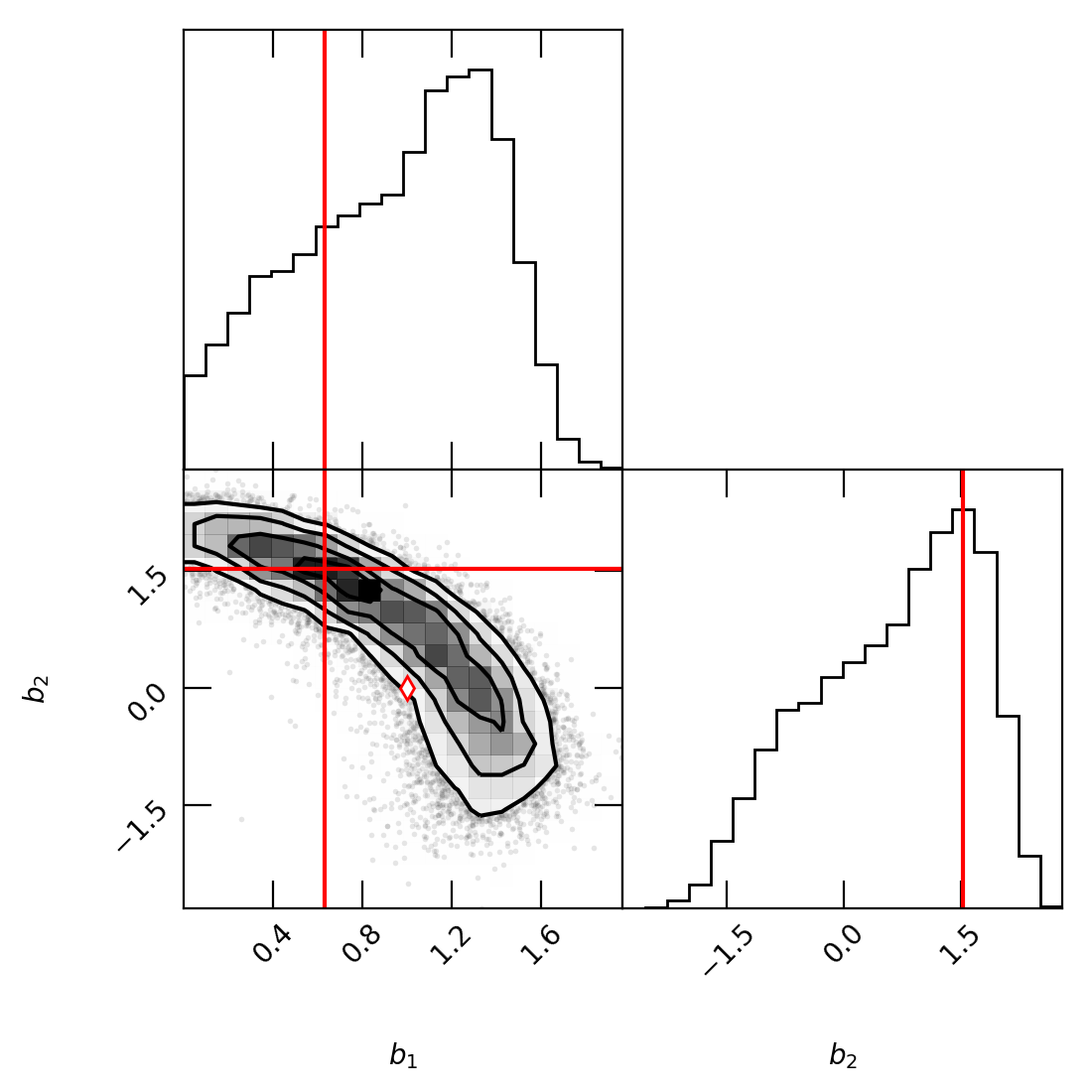}
    \caption{Posterior and marginalized distributions of $b_1$ and $b_2$ obtained through a joint Markov Chain Monte Carlo sampling of the auto power spectrum of binary black hole mergers, and cross power spectrum between mergers and matter computed at redshift $z=0$. The red lines indicate the maximum a posteriori estimate of the parameters, while the red diamond indicates the initial guess.}
    \label{fig:corner_bbh_z0}
\end{figure}

\begin{figure}
    \centering
    \includegraphics[width=0.4\textwidth]{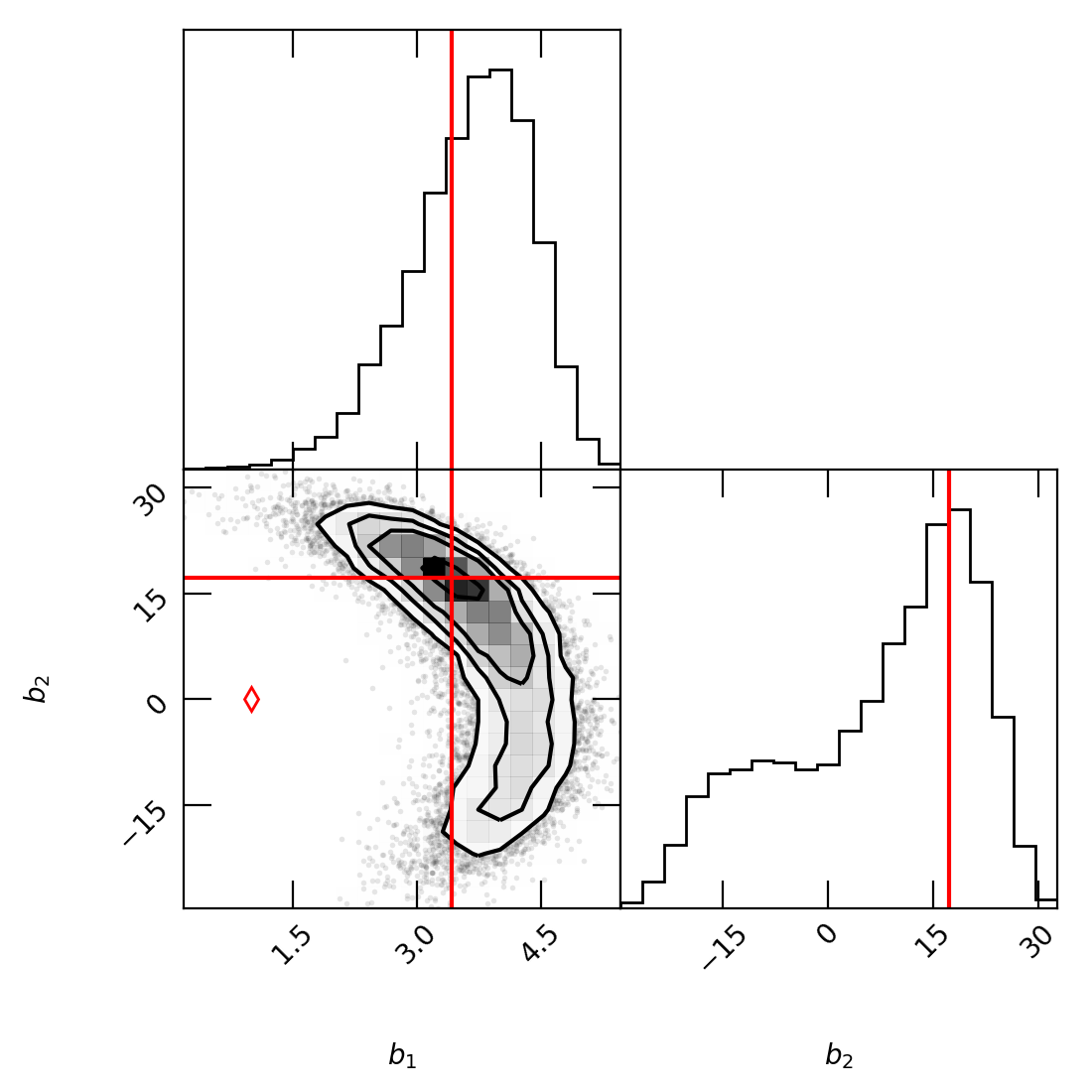}
    \caption{Same as \ref{fig:corner_bbh_z0} at $z=6$.}
    \label{fig:corner_bbh_z6}
\end{figure}

%%%%%%%%%%%%%%%%%%%%%%%%%%%%%%%%%%%%%%%%%%%%%%%%%%

% Don't change these lines
\bsp	% typesetting comment
\label{lastpage}
\end{document}